\documentclass[10pt,journal,epsfig]{IEEEtran}
\usepackage[dvips]{graphicx}
\usepackage{graphicx}
\usepackage{amssymb}
\usepackage{cite}
\usepackage{subfigure}
\usepackage{multirow,boxedminipage}
\usepackage{slashbox,booktabs}
\usepackage{bm,comment}
\usepackage{amsmath}
\newcounter{MYtempeqncnt}

\usepackage{subeqnarray}
\usepackage{algorithm}
\usepackage{algpseudocode}
\usepackage{cases}
\usepackage[Symbol]{upgreek}
\makeatletter

\newcommand{\Rmnum}[1]{\expandafter\@slowromancap\romannumeral #1@}

\makeatother

\newtheorem{property}{Property}
\newtheorem{proposition}{Proposition}
\newtheorem{remark}{Remark}

\newtheorem{lemma}{Lemma}
\newtheorem{corollary}{Corollary}

\newtheorem{condition}{Condition}

\begin{document}
\IEEEoverridecommandlockouts
\title{Artificial Noise-Aided Biobjective Transmitter Optimization for Service Integration in Multi-User MIMO Gaussian Broadcast Channel}
\author{Weidong Mei, Zhi Chen, \IEEEmembership{Senior Member, IEEE}, Jun Fang \IEEEmembership{Member, IEEE} and Shaoqian Li \IEEEmembership{Fellow, IEEE}
\thanks{This work was supported in part by the National Natural Science Foundation of China under Grants 61631004 and 61571089.}
\thanks{The authors are with National Key Laboratory of Science and Technology on Communications, University of Electronic Science and Technology of China, Chengdu (611731), China (e-mails: mwduestc@gmail.com; chenzhi@uestc.edu.cn; JunFang@uestc.edu.cn; lsq@uestc.edu.cn)
}}
\maketitle

\begin{abstract}
This paper considers an artificial noise (AN)-aided transmit design for multi-user MIMO systems with integrated services. Specifically, two sorts of service messages are combined and served simultaneously: one multicast message intended for all receivers and one confidential message intended for only one receiver and required to be perfectly secure from other unauthorized receivers. Our interest lies in the joint design of input covariances of the multicast message, confidential message and artificial noise (AN), such that the achievable secrecy rate and multicast rate are simultaneously maximized. This problem is identified as a secrecy rate region maximization (SRRM) problem in the context of physical-layer service integration. Since this bi-objective optimization problem is inherently complex to solve, we put forward two different scalarization methods to convert it into a scalar optimization problem. First, we propose to prefix the multicast rate as a constant, and accordingly, the primal biobjective problem is converted into a secrecy rate maximization (SRM) problem with quality of multicast service (QoMS) constraint. By varying the constant, we can obtain different Pareto optimal points. The resulting SRM problem can be iteratively solved via a provably convergent difference-of-concave (DC) algorithm. In the second method, we aim to maximize the weighted sum of the secrecy rate and the multicast rate. Through varying the weighted vector, one can also obtain different Pareto optimal points. We show that this weighted sum rate maximization (WSRM) problem can be recast into a primal decomposable form, which is amenable to alternating optimization (AO). Then we compare these two scalarization methods in terms of their overall performance and computational complexity via theoretical analysis as well as numerical simulation, based on which new insights can be drawn.
\end{abstract}

\section{Introduction}
\subsection{Background}
Recently, \emph{physical-layer service integration} (PHY-SI), a technique of combining multicast service and confidential service into one integrated service for one-time transmission at the physical layer, has received much attention in wireless communications. For one thing, PHY-SI caters to the demand for high transmission rate and secure communication, which has been identified as the key targets that need to be effectively addressed by fifth generation (5G) wireless systems \cite{andrew2014what}. Besides, compared with the conventional upper-layer-based approach, PHY-SI enables coexisting services to share the same resources by solely exploiting the physical characteristics of wireless channels, thereby significantly increasing the spectral efficiency. This property makes PHY-SI a prominent approach to satisfy the ever-increasing need for radio spectrum. The technique of PHY-SI could also find a wide range of applications in the commercial and military areas. For example, many commercial applications, e.g., advertisement, digital television, Internet telephony, and so on, are supposed to provide personalized service customization. As a consequence, confidential service and public service are collectively provided to satisfy the demand of different user groups. In battlefield scenarios, it is essential to propagate commands with different security levels to the frontline. The public information should be distributed to all soldiers, while the confidential information can only be accessed by specific soldiers. Such emerging applications lead to a crucial problem: \emph{how to establish the security of confidential service while not compromising the quality of public service}?

\subsection{Related Works}
Let us first have a very brief review on physical-layer security, a technique that lays foundation for the research on PHY-SI. The broadcast nature of wireless medium makes privacy an inherent concern. Physical layer security technique is playing an increasingly important role in wireless communication recently. It can secure communications information-theoretically at the physical layer without using secret keys whose distribution or management may become difficult in e.g., ad-hoc wireless networks. Different strategies against eavesdropping have been developed with various levels of channel state information (CSI) available to the transmitter (see the comprehensive overview in \cite{shiu2011physical,he2013wireless,hong2013enhancing,mukherjee2014principles,liu2016physical}). Liu and Poor first coined the term \emph{confidential broadcasting} in \cite{liu2009secrecy,liu2010multiple} and established the corresponding secrecy capacity region. In confidential broadcasting, a transmitter broadcasts multiple confidential messages to all receivers. Each confidential message is intended for one specified receiver but required to be perfectly secret from the others. Some efforts have been made in e.g., \cite{fakoorian2013on,park2016weighted} to maximize the sum secrecy rate in the scenario of confidential broadcasting.

The study of PHY-SI can be traced back to Csisz{\'a}r and K{\"o}rner's seminar work in \cite{csiszar1978broadcast}. In the basic model of PHY-SI, a transmitter sends a common message to two receivers, and simultaneously, sends a confidential message intended only for one receiver and kept perfectly secure from the other one. Under discrete memoryless broadcast channel (DMBC) setup, Csisz{\'a}r and K{\"o}rner gave a closed-form expression of the maximum rate region that can be applied reliably under the secrecy constraint (i.e., the secrecy capacity region). In recent years, this kind of approach has gained renewed interest, especially that in various multi-antenna scenarios, such as Gaussian broadcast channels \cite{Hung2010Multiple, ekrem2012capacity, liu2010mimo, liu2013new}, and bidirectional relay channels \cite{wyrembelski2011service, Wyrembelski2012Physical}. In \cite{Hung2010Multiple}, the authors extended the results in \cite{csiszar1978broadcast} to a general MIMO Gaussian case by adopting the channel-enhancement argument. Further, the works \cite{ekrem2012capacity, liu2010mimo} considered the case with two confidential messages intended for two different receivers. The resulting secrecy capacity region is proved to be attainable by combining the secret dirty-paper coding (S-DPC) with Gaussian superposition coding. Furthermore, in \cite{wyrembelski2011service} and \cite{Wyrembelski2012Physical}, Wyrembelski and Boche amalgamated broadcast service, multicast service and confidential service in bidirectional relay networks, in which a relay adds an additional multicast message for all nodes and a confidential message for only one node besides establishing the conventional bidirectional communication. Nonetheless, the main goal of the aforementioned papers is just to obtain capacity results or to characterize coding strategies that lead to certain rate regions\cite{Schaefer2014Physical}. For implementation efficiency, it is also important to treat physical layer service integration from a signal processing point of view. In particular, optimal or complexity-efficient transmit strategies have to be characterized, so that the achieved performance could reach/approach the boundary of the secrecy capacity region. Such strategies are usually given by optimization problems, which generally turn out to be nonconvex. Along with this comes the fact that most works on PHY-SI end once a certain characterization of a rate region is derived.

Recently, to fill in the gap between the previous information-theoretic results and practical implementation, there is growing interest in analyzing PHY-SI from a signal processing point of view. In \cite{Hung2010Multiple}, the authors proposed a \emph{re-parameterizing} method to devise transmit strategies for achieving the secrecy boundary performance. However, this method is only applicable to a very simple two-user MISO scenario. Besides, it involves solving a sequence of convex feasibility problems, which is computationally expensive. To improve it, the work \cite{mei2016secrecy} proposed a \emph{quality-of-service (QoS)-based} method to seek the boundary-achieving transmit strategies. Its basic idea is to establish the tradeoff between the secrecy rate and the multicast rate by maximizing the secrecy rate while ensuring the multicast rate above a given threshold. This method is demonstrated as effective in characterizing the secrecy boundary, and thus triggered research endeavors on extending the result to a more general and realistic setting. Notable results include the extension to the multi-user \cite{mei1512} and imperfect CSI \cite{mei2016robust, mei2016artificial} settings. Even so, relatively less work focussed on the MIMO channel setup, due to the intractability of the associated optimization problems. In \cite{mei2016GSVD}, the authors circumvented the intractability by proposing a generalized singular value decomposition (GSVD) based transmission scheme. Using GSVD, multicast message and confidential message can be perfectly decoupled and the resulting problem is easier to handle. However, this result is not applicable to the general multi-user MIMO case. In addition, it is also interesting to incorporate artificial noise (AN) into consideration, as such technique has been shown to be effective in enhancing transmission security\cite{li2013transmit,li2013spatially,chu2015robust,chu2016secrecy,zheng2015multi}. Specifically, the authors in \cite{li2013transmit}, \cite{li2013spatially,chu2015robust,chu2016secrecy} and \cite{zheng2015multi} respectively showed that AN is of paramount importance to physical-layer security when there exist multiple eavesdroppers in the network, when the CSI of eavesdropper(s) is imperfectly known at the transmitter, and/or when eavesdroppers are randomly located in the network.

\subsection{Main Contributions}
In this paper, we delve into the AN-aided transmit precoding design in PHY-SI under a general multi-user MIMO case. Specifically, two sorts of service messages are combined and promulgated at the same time: a multicast message intended for all receivers, and a confidential message intended for merely one authorized receiver. The confidential message must be kept perfectly secure from all other unauthorized receivers. Meanwhile, AN is employed to degrade the potential eavesdropping of the unauthorized receivers. This paper aims to jointly optimize the input covariance matrices of the multicast message, confidential message and AN, to maximize the achievable secrecy and multicast rates simultaneously, or equivalently, to maximize the achievable secrecy rate region. This secrecy rate region maximization (SRRM) problem turns out to be a biobjective optimization problem. Since the re-parameterizing method is invalid in a general MIMO case, we develop two scalarization methods to convert it into an easier-to-handle scalar version for characterizing its Pareto boundary.
\begin{enumerate}
  \item In the first method, we propose to fix the multicast rate as a constant. Through varying the value of the constant, this method could yield different secrecy boundary points. Since the Pareto optimal points must reside on the boundary of the achievable rate region, this method is bound to provide a complete set of the Pareto optimal points. Though the resultant secrecy rate maximization (SRM) problem is nonconvex by nature, we show this problem actually falls into the context of difference-of-concave (DC) programming \cite{NIPS2009}. Hence, it can be handled by classical DC algorithm with convergence guarantee.
  \item As for the second method, a weighted sum-based scalarization is introduced. The crux of this scalarization method is to optimize the weighted sum of the two objectives with different weight vectors. By varying the weight vector, this method gives rise to different Pareto optimal solution. To solve this weighted sum rate maximization (WSRM) problem, we reveal its hidden decomposability by recasting it as an equivalent form amenable to alternating optimization (AO). AO algorithm is naturally employed to solve the WSRM problem. It can be proved that this AO algorithm must converge to one stationary point of the WSRM problem.
  \item It is particularly worth mentioning that though the DC and AO algorithms have been applied to address the issue of physical-layer security before in e.g., \cite{li2013transmit,fang2016precoding,chu2015secrecy}, none of these works considered integrating an additional multicast message. Our paper is an initial attempt to study the application of DC and AO to the emerging PHY-SI system, which turns out to be a harder task than its counterpart in physical-layer security due to the coexisting multicast service.
\end{enumerate}

Then we compare these two sorts of scalarization methods in terms of their overall performance and computational complexity. The comparison results reveal that the first method is more efficacious in finding all Pareto optimal points than the second one. The advantage of the second method lies in its problem structure, which provides the service provider a solution to maximizing the overall revenue. Besides, we show that the DC algorithm is more time-efficient at low transmit power than the AO algorithm. Interestingly, the numerical results indicate that at high transmit power, the AO algorithm becomes the more time-efficient one.

\subsection{Organization and Notations}
This paper is organized as follows. Section \Rmnum{2} provides the system model description and problem formulation. The optimization aspects of our formulated problems are addressed in Section \Rmnum{3} and \Rmnum{4}, corresponding to the first and the second scalarization methods, respectively. The comparison results are given in Sections \Rmnum{5}. Section \Rmnum{6} presents simulation results to show the efficacy of our proposed methods. Finally, conclusions are drawn in Section \Rmnum{7}.

The notation of this paper is as follows. Bold symbols in capital letter and small letter denote matrices and vectors, respectively. ${{(\cdot)}^{H}}$, $\rm{rank}(\cdot)$ and $\text{Tr}(\cdot )$ represent conjugate transpose, rank and trace of a matrix, respectively. ${\mathbb{R}}_{+}$ and ${\mathbb{H}}_{+}^n$ denote the set of nonnegative real numbers and of $n$-by-$n$ Hermitian positive semidefinite (PSD) matrices. The $n \times n$ identity matrix is denoted by ${\mathbf{I}}_n$. $\mathbf{x}\sim \mathcal{CN}(\mathbf{\mu },\mathbf{\Omega })$ denotes that \textbf{x} is a complex circular Gaussian random vector with mean $\mathbf{\mu}$ and covariance $\mathbf{\Omega}$. $\mathbf{A}\succeq \mathbf{0}$ $(\mathbf{A}\succ \mathbf{0})$ implies that $\mathbf{A}$ is a Hermitian positive semidefinite (definite) matrix. ${\left\| \cdot \right\|}$ represents the vector Euclidean norm. $K$ represents a proper cone, and $K^*$ represents a dual cone associated with $K$.

\section{System Model and Problem Formulation}
We consider the downlink of a multiuser system in which a multi-antenna transmitter serves $K$ receivers, and each receiver is equipped with multiple antennas. Assume that all receivers have ordered the multicast service and receiver 1 further ordered the confidential service\footnote{In this paper, we assume that only one receiver orders the confidential service within a single time slot. In practice, this assumption is valid under the case where the confidential service is provided to all receivers in a \emph{round-robin} manner, i.e., the time slots are assigned to each subscriber of the confidential service in equal portions and in circular order.}. To enhance the security performance, the transmitter utilizes a fraction of its transmit power to send artificially generated noise to interfere the unauthorized receivers (eavesdroppers), i.e., receiver 2 to receiver $K$. We assume in this paper that all receivers are static and that all the communication links undergo slow frequency-flat fading.

\begin{remark}
In this paper, we assume that only one receiver orders the confidential service within a single time slot. In practice, this assumption is valid under the cases where the confidential service is provided to all receivers in a \emph{round-robin} manner to strengthen the security of confidential messages and to reduce the operational complexity at the transmitter.
\end{remark}

The received signal at receiver $k$ is modeled as
\begin{equation}\label{yk}
  {{\mathbf{y}}_k} = \;{{\bf{H}}_k}{\bf{x}} + {{\mathbf{z}}_k}, k=1,2,\cdots,K
\end{equation}
where ${{\mathbf{H}}_k}\in {{\mathbb{C}}^{{{N}_{r,k} \times {N}_{t}}}}$ is the channel response between the transmitter and receiver $k$; and ${N}_{t}$ and ${N}_{r,k}$ are the number of transmit antennas employed by the transmitter and $k$th receiver, respectively. ${{\mathbf{z}}_k}$ is independent identically distributed (i.i.d.) complex Gaussian noise with zero mean and unit variance. ${{\mathbf{x}}}\in {{\mathbb{C}}^{{{N}_{t}}}}$ is the coded transmit message, which consists of three independent components, i.e.,
\begin{equation}\label{x3c}
  {\bf{x\;}}\; = \;{{\bf{x}}_0} + \;{{\bf{x}}_c} + \;{{\bf{x}}_a},
\end{equation}
where ${\bf{x}}_{0}$ is the multicast message intended for all receivers, ${\bf{x}}_{c}$ is the confidential message intended for receiver 1, and ${\bf{x}}_{a}$ is the artificial noise. We assume $\mathbf{x}_{0} \sim \mathcal{CN}(\mathbf{0},\mathbf{Q}_0)$, $\mathbf{x}_{c} \sim \mathcal{CN}(\mathbf{0},\mathbf{Q}_c)$ \cite{Hung2010Multiple}, where $\mathbf{Q}_0$ and $\mathbf{Q}_c$ are the transmit covariance matrices. The AN ${\bf{x}}_{a}$ follows a distribution $\mathbf{x}_{a} \sim \mathcal{CN}(\mathbf{0},\mathbf{Q}_a)$, where $\mathbf{Q}_a$ is the AN covariance. The CSI on all links is assumed to be perfectly known at the corresponding transmitter and receivers in that all receivers have to register in the network for subscribing the multicast service. In practice, the CSI at the receivers can be obtained from the channel estimation of the downlink pilots. CSI at the transmitter can be acquired via uplink channel estimation in time division duplex (TDD) systems. The design of a high-quality channel estimation scheme is beyond the scope of this paper. Note that the full CSI assumption is commonly adopted in the area of physical layer security/multicasting, especially in MIMO channels \cite{li2013transmit,fang2016precoding,yang2013optimal,park2016weighted,wu2013physical,zhu2012precoder,lee2013a,du2013optimum}.

For ease of exposition, let us define ${\cal K} \buildrel \Delta \over = \{1,2,...,K\}$ and ${{\cal K}_e} \buildrel \Delta \over = {\cal K}/\{ 1\} $, which denote the indices of all receivers and of all unauthorized receivers, respectively. Denote $R_0$ and $R_c$ as the achievable rates associated with the multicast and confidential messages, respectively. Then an achievable secrecy rate region $R_s({\left\{ {{\bf{H}}_k} \right\}_{k \in {\cal{K}}}},P)$ is given as the set of nonnegative rate pairs $(R_0,R_c)$ satisfying \cite{Hung2010Multiple}
\begin{equation}\label{Region1}
\begin{split}
&{R_0} \le \mathop {\min }\limits_{k \in {\cal K}} C_{m,k}({{\bf{Q}}_0},{{\bf{Q}}_c},{{\bf{Q}}_a})\\
&{R_c} \le  C_b({{\bf{Q}}_c},{{\bf{Q}}_a}) - \mathop {\max}\limits_{k \in {{\cal K}_e}} C_{e,k}({{\bf{Q}}_c},{{\bf{Q}}_a}),
\end{split}
\end{equation}
where
\begin{subequations}\label{Region2}
\begin{align}
\nonumber &C_{m,k}({{\bf{Q}}_0},{{\bf{Q}}_c},{{\bf{Q}}_a})=\\
&\qquad{\log | {{\bf{I}} + {{({\bf{I}} + {{\bf{H}}_k}({{\bf{Q}}_c} + {{\bf{Q}}_a}){\bf{H}}_k^H)}^{ - 1}}{{\bf{H}}_k}{{\bf{Q}}_0}{\bf{H}}_k^H} |},\\
&C_b({{\bf{Q}}_c},{{\bf{Q}}_a}) = \log | {{\bf{I}} + {{({\bf{I}} + {{\bf{H}}_1}{{\bf{Q}}_a}{\bf{H}}_1^H)}^{ - 1}}{{\bf{H}}_1}{{\bf{Q}}_c}{\bf{H}}_1^H} |,\\
&C_{e,k}({{\bf{Q}}_c},{{\bf{Q}}_a})= \log | {{\bf{I}} + {{({\bf{I}} + {{\bf{H}}_k}{{\bf{Q}}_a}{\bf{H}}_k^H)}^{ - 1}}{{\bf{H}}_k}{{\bf{Q}}_c}{\bf{H}}_k^H} |,
\end{align}
\end{subequations}
and $\text{Tr}(\mathbf{Q}_0+\mathbf{Q}_c+\mathbf{Q}_a) \le P$ with $P$ being total transmit power budget at the transmitter.

The secrecy rate region (\ref{Region1}) implies that all receivers first decode their common multicast message by treating the confidential message as noise, and then receiver 1 acquires a clean link for the transmission of its exclusive confidential message, where there is no interference from the multicast message. This can be achieved by utilizing the encoding schemes proposed in \cite{csiszar1978broadcast}.

To maximize this achievable secrecy rate region, our goal is to find the boundary-achieving $\mathbf{Q}_0$, $\mathbf{Q}_a$ and $\mathbf{Q}_c$, which is also known as Pareto optimal solutions to this SRRM problem. Specifically, with perfect CSI available at the transmitter, we must first solve the following optimization problem, which is a biobjective maximization problem with cone $K=K^*={\mathbb {R}}_ + ^2$.
\begin{subequations}\label{op1}
\begin{align}
\nonumber&\mathop{\max}\limits_{{{\bf{Q}}_0},{{\bf{Q}}_c},{{\bf{Q}}_a}} \left({\text{w.r.t.}}\; {\mathbb {R}}_ + ^2 \right)\; ( {\mathop {\min }\limits_{k \in {\cal K}} C_{m,k},C_b - \mathop {\max}\limits_{k \in {{\cal K}_e}} C_{e,k}} )\\
\text{s.t.}\quad&\text{Tr}({{\bf{Q}}_0} + {{\bf{Q}}_c} + {\bf{Q}}_a) \le P,\label{op1a}\\
&{{\bf{Q}}_0} \succeq {\bf{0}}, {{\bf{Q}}_c} \succeq {\bf{0}}, {\bf{Q}}_a \succeq {\bf{0}},\label{op1b}
\end{align}
\end{subequations}
where, with a slight abuse of notations but for notational simplicity, the explicit dependence of $C_{m,k}$, $C_b$ and $C_{e,k}$ on $({{\bf{Q}}_0},{{\bf{Q}}_c},{{\bf{Q}}_a})$ is omitted. Since the SRRM problem is a biobjective maximization problem, it is necessary to harness some method of scalarization to convert it into an easier-to-handle scalar version.

\begin{remark}
It is also viable to consider the scenario where all receivers order the confidential service and all confidential messages are propagated concurrently by the transmitter, i.e., the integration of multicasting and confidential broadcasting. The merit of this scheme lies in its higher spectral efficiency and low latency. However, this comes at the expense of much higher operational complexity at the transmitter, especially when the number of users increases. Thus, our considered PHY-SI scheme is particularly desired in delay-tolerant applications or when the transmitter possesses limited computational capacity for security-related computations.
\end{remark}

\section{A DC-Based Approach to the SRRM Problem}
In this section, we develop our first scalarization method to solve (\ref{op1}). The basic problem formulation is a secrecy rate maximization (SRM) with imposed quality of multicast service (QoMS) constraints.

\subsection{Scalarization}
In particular, our method is to move the multicast rate maximization part to the constraint, i.e., we fix at the time being the multicast rate as a constant $\tau_{ms} \ge 0$. As a result, the biobjective SRRM problem (\ref{op1}) will be degraded into a scalar maximization problem, which is shown in (\ref{op2}).
\begin{subequations}\label{op2}
\begin{align}
\nonumber R&(\tau _{ms}) =\mathop{\max}\limits_{{{\bf{Q}}_0},{{\bf{Q}}_c},{{\bf{Q}}_a}} C_b({{\bf{Q}}_c},{{\bf{Q}}_a})-\mathop {\max }\limits_{k \in {\cal K}_e}C_{e,k}({{\bf{Q}}_c},{{\bf{Q}}_a})\\
\text{s.t.}\; &\mathop {\min }\limits_{k \in {\cal K}}C_{m,k}({{\bf{Q}}_0},{{\bf{Q}}_c},{{\bf{Q}}_a}) = {\tau_{ms}}, \label{op2a}\\
&\text{Tr}({{\bf{Q}}_0} + {{\bf{Q}}_c}+{{\bf{Q}}_a}) \le P, \\
&{{\bf{Q}}_0} \succeq {\bf{0}}, {{\bf{Q}}_a} \succeq {\bf{0}}, {{\bf{Q}}_c} \succeq {\bf{0}}.
\end{align}
\end{subequations}
In (\ref{op2}), $R(\tau _{ms})$ is the optimal objective value, $\tau _{ms}$ can be interpreted as the preset requirement on the multicast rate, and accordingly, the constraint (\ref{op2a}) can be interpreted as a QoMS constraint. To guarantee the feasibility of problem (\ref{op2}), $\tau _{ms}$ cannot exceed a threshold $\tau _{\max}$ given by
\begin{equation}\label{max_tau}
{\tau _{\max }} = \mathop {\max }\limits_{{{\bf{Q}}_0} \succeq {\bf{0}}, \text{Tr}({{\bf{Q}}_0}) \le P} \mathop {\min }\limits_{k \in {\cal K}} \log \left| {{\bf{I}} + {{\bf{H}}_k}{{\bf{Q}}_0}{\bf{H}}_k^H} \right|.
\end{equation}
The value of $\tau _{\max}$ can be numerically obtained by solving (\ref{max_tau}) via the convex optimization solver \texttt{CVX} \cite{Boyd2011CVX}.

This sort of problem formulation, in fact, enables us to find one boundary point $(\tau_{ms},R(\tau _{ms}))$ of the secrecy rate region $R_s({\left\{ {{\bf{H}}_k} \right\}_{k \in {\cal{K}}}},P)$ by solving (\ref{op2}). All boundary points of $R_s({\left\{ {{\bf{H}}_k} \right\}_{k \in {\cal{K}}}},P)$ can be found if we traverse all possible ${\tau_{ms}}$'s lying within $[0,{\tau _{\max }}]$ and store the corresponding optimal objective values. Since the Pareto optimal solution to (\ref{op1}) must reside on the boundary of $R_s({\left\{ {{\bf{H}}_k} \right\}_{k \in {\cal{K}}}},P)$, i.e., the Pareto optimal set of (\ref{op1}) is a \emph{subset} of the boundary set of $R_s({\left\{ {{\bf{H}}_k} \right\}_{k \in {\cal{K}}}},P)$, all Pareto optimal solution to (\ref{op2}) can also be found by this means.

However, problem (\ref{op2}) is nonconvex. Especially, the determinant equality constraint (\ref{op2a}) is very difficult to handle. To circumvent this difficulty, we pay our attention to the following relaxed problem of (\ref{op2}), in which the equality constraint (\ref{op2a}) is replaced by the inequality constraint (\ref{relax.a}).
\begin{subequations}\label{relax}
\begin{align}
\nonumber \tilde R&(\tau _{ms}) =\mathop{\max}\limits_{{{\bf{Q}}_0},{{\bf{Q}}_c},{{\bf{Q}}_a}} C_b({{\bf{Q}}_c},{{\bf{Q}}_a})-\mathop {\max }\limits_{k \in {\cal K}_e}C_{e,k}({{\bf{Q}}_c},{{\bf{Q}}_a})\\
\text{s.t.}\; &\mathop {\min }\limits_{k \in {\cal K}}C_{m,k}({{\bf{Q}}_0},{{\bf{Q}}_c},{{\bf{Q}}_a}) \ge {\tau_{ms}}, \label{relax.a}\\
&\text{Tr}({{\bf{Q}}_0} + {{\bf{Q}}_c}+{{\bf{Q}}_a}) \le P, \label{relax.b}\\
&{{\bf{Q}}_0} \succeq {\bf{0}}, {{\bf{Q}}_a} \succeq {\bf{0}}, {{\bf{Q}}_c} \succeq {\bf{0}}.
\end{align}
\end{subequations}
Apparently, any optimal solution to (\ref{op2}) is feasible to (\ref{relax}) in the sense that replacing (\ref{op2a}) with (\ref{relax.a}) yields a larger feasible solution set. Hence, problem (\ref{relax}) has $R(\tau _{ms}) \le \tilde R(\tau _{ms})$ in general. Interestingly, we show that $R(\tau _{ms}) = \tilde R(\tau _{ms})$ can always be achieved without loss of optimality to (\ref{relax}).
\begin{lemma}\label{equivalent}
Problem (\ref{relax}) is a tight relaxation to problem (\ref{op2}). In other words, the rate pair $({\tau_{ms}},\tilde R(\tau _{ms}))$ must be a boundary point of $R_s({\left\{ {{\bf{H}}_k} \right\}_{k \in {\cal{K}}}},P)$.
\end{lemma}
\begin{IEEEproof}
The proof can be easily accomplished by construction. Suppose that the constraint (\ref{relax.a}) is satisfied with strict inequality, we can always multiply ${{\bf{Q}}_0}$ by a scalar $\nu\; (\nu < 1)$ to make (\ref{relax.a}) active, yet without decreasing the objective value of (\ref{relax}) and violating the total power constraint (\ref{relax.b}). This fact implies that there always exists an optimal solution to (\ref{relax}) such that the constraint (\ref{relax.a}) is satisfied with equality, and thus, accomplishes the proof.
\end{IEEEproof}

Lemma \ref{equivalent} implies that problem (\ref{relax}) admits an optimal $({{\bf{Q}}^*_0},{{\bf{Q}}^*_c},{{\bf{Q}}^*_a})$ with $\mathop {\min }\limits_{k \in {\cal K}}C_{m,k}({{\bf{Q}}^*_0},{{\bf{Q}}^*_c},{{\bf{Q}}^*_a})={\tau_{ms}}$. Hence, $({{\bf{Q}}^*_0},{{\bf{Q}}^*_c},{{\bf{Q}}^*_a})$ is also optimal to (\ref{op2}). The proof of Lemma \ref{equivalent} reveals that such an optimal $({{\bf{Q}}^*_0},{{\bf{Q}}^*_c},{{\bf{Q}}^*_a})$ can always be constructed algorithmically based on the following procedures:

\begin{corollary}
Suppose that $({{\bf{Q}}^*_0},{{\bf{Q}}^*_c},{{\bf{Q}}^*_a})$ is an optimal solution returned by solving problem (\ref{relax}). If $\mathop {\min }\limits_{k \in {\cal K}}C_{m,k}({{\bf{Q}}^*_0},{{\bf{Q}}^*_c},{{\bf{Q}}^*_a})={\tau_{ms}}$, then output $({{\bf{Q}}^*_0},{{\bf{Q}}^*_c},{{\bf{Q}}^*_a})$ as an optimal solution of problem (\ref{op2}). Otherwise, solve the following equation with regard to $\nu$, i.e., $\mathop {\min }\limits_{k \in {\cal K}}C_{m,k}(\nu{{\bf{Q}}^*_0},{{\bf{Q}}^*_c},{{\bf{Q}}^*_a})={\tau_{ms}}$, via bisection search within the unit interval $[0,1]$, and output $(\nu{{\bf{Q}}^*_0},{{\bf{Q}}^*_c},{{\bf{Q}}^*_a})$ as an optimal solution of problem (\ref{op2}).
\end{corollary}

Next, we will point out two special cases, under which problem (\ref{relax}) is \emph{equivalent} to problem (\ref{op2}); or equivalently, any optimal solution to (\ref{relax}) is achieved with constraint (\ref{relax.a}) active. This is described in the following proposition.
\begin{proposition}\label{P1}
Suppose that the system configurations satisfy either one of the following conditions:
\begin{condition}\label{C1}
The number of antennas at the transmitter is larger than that at the authorized receiver, i.e., $N_t > N_{r,1}$.
\end{condition}
\begin{condition}\label{C2}
The number of antennas at the transmitter is larger than the sum of the antenna number at the unauthorized receivers, i.e., $N_t > \sum\nolimits_{k \in {{\cal K}_e}} N_{r,k}$.
\end{condition}
Then the rate pair $({\tau_{ms}},\tilde R(\tau _{ms}))$ must be a \emph{Pareto optimal} point of (\ref{op1}), and all Pareto optimal points of (\ref{op1}) can be obtained by solving (\ref{op2}) with different $\tau_{ms}$'s lying within the interval $[0,\tau_{\max}]$.
\end{proposition}
\begin{IEEEproof}
The proof can be found in Appendix \ref{DC_Appendix}.
\end{IEEEproof}

\begin{remark}
Proposition \ref{P1} bridges the Pareto optimal points of (\ref{op1}) to the boundary points of $C_s({\bf{H}}_1,{\bf{H}}_2,P)$. When either Condition \ref{C1} or Condition \ref{C2} is satisfied, all Pareto optimal points of (\ref{op1}) are also the boundary points of $R_s({\left\{ {{\bf{H}}_k} \right\}_{k \in {\cal{K}}}},P)$, and vice versa.
\end{remark}

\subsection{DC Iterative Algorithm}
We now focus on solving the relaxed problem (\ref{relax}) derived in the last subsection. Problem (\ref{relax}) still remains nonconvex due to its objective function and constraint (\ref{relax.a}). To deal with it, we first equivalently transform it into its epigraph form by introducing a slack variable $\eta$, i.e.,
\begin{subequations}\label{op3}
\begin{align}
\nonumber R(\tau _{ms})& =\mathop{\max}\limits_{{{\bf{Q}}_0},{{\bf{Q}}_c},{{\bf{Q}}_a},\eta} C_b({{\bf{Q}}_c},{{\bf{Q}}_a})-\eta\\
\text{s.t.}\; &C_{e,k}({{\bf{Q}}_c},{{\bf{Q}}_a}) \le \eta, \forall k \in {\cal{K}}_e\\
&C_{m,k}({{\bf{Q}}_0},{{\bf{Q}}_c},{{\bf{Q}}_a}) \ge {\tau_{ms}}, \forall k \in {\cal{K}}\label{op3a}\\
&\text{Tr}({{\bf{Q}}_0} + {{\bf{Q}}_c}+{{\bf{Q}}_a}) \le P, \\
&{{\bf{Q}}_0} \succeq {\bf{0}}, {{\bf{Q}}_a} \succeq {\bf{0}}, {{\bf{Q}}_c} \succeq {\bf{0}}.
\end{align}
\end{subequations}

Next, we will show that problem (\ref{op3}) constitutes a DC-type programming problem, which can be iteratively solved by employing the DC algorithm.

To begin with, we reformulate the capacity function $C_b({{\bf{Q}}_c},{{\bf{Q}}_a})$, $C_{e,k}({{\bf{Q}}_c},{{\bf{Q}}_a})$ and $C_{m,k}({{\bf{Q}}_0},{{\bf{Q}}_c},{{\bf{Q}}_a})$ into a DC-type form, given by
\begin{align}\label{DC1}
&C_b({{\bf{Q}}_c},{{\bf{Q}}_a})= {\phi _1}({{\bf{Q}}_c},{{\bf{Q}}_a}) - {\varphi _1}({{\bf{Q}}_a}),\nonumber\\
&C_{e,k}({{\bf{Q}}_c},{{\bf{Q}}_a})={\phi _k}({{\bf{Q}}_c},{{\bf{Q}}_a}) - {\varphi _k}({{\bf{Q}}_a}),\forall k \in {\cal{K}}_e \\
&C_{m,k}({{\bf{Q}}_0},{{\bf{Q}}_c},{{\bf{Q}}_a})={\eta _k}({{\bf{Q}}_0},{{\bf{Q}}_c},{{\bf{Q}}_a}) - {\phi _k}({{\bf{Q}}_c},{{\bf{Q}}_a}),\forall k \in {\cal{K}}\nonumber
\end{align}
in which we define
\begin{align}
&{\phi _k}({{\bf{Q}}_c},{{\bf{Q}}_a}) = \log \left| {{\bf{I}} + {{\bf{H}}_k}({{\bf{Q}}_c} + {{\bf{Q}}_a}){\bf{H}}_k^H} \right|,\forall k \in {\cal{K}}\nonumber\\
&{\varphi _k}({{\bf{Q}}_a}) = \log \left| {{\bf{I}} + {{\bf{H}}_k}{{\bf{Q}}_a}{\bf{H}}_k^H} \right|,\forall k \in {\cal{K}}\label{DC2}\\
&{\eta _k}({{\bf{Q}}_0},{{\bf{Q}}_c},{{\bf{Q}}_a}) = \log \left| {{\bf{I}} + {{\bf{H}}_1}({{\bf{Q}}_c} + {{\bf{Q}}_a} + {{\bf{Q}}_0}){\bf{H}}_1^H} \right|, \forall k \in {\cal{K}}.\nonumber
\end{align}

Substituting (\ref{DC1}) into problem (\ref{op3}), we obtain
\begin{subequations}\label{op4}
\begin{align}
\nonumber R&(\tau _{ms}) =\mathop{\max}\limits_{{{\bf{Q}}_0},{{\bf{Q}}_c},{{\bf{Q}}_a},\eta} {\phi _1}({{\bf{Q}}_c},{{\bf{Q}}_a}) - {\varphi _1}({{\bf{Q}}_a})-\eta\\
\text{s.t.}\; &{\varphi _k}({{\bf{Q}}_a})-{\phi _k}({{\bf{Q}}_c},{{\bf{Q}}_a})+ \eta \ge 0, \forall k \in {\cal{K}}_e\label{op4a}\\
&{\eta _k}({{\bf{Q}}_0},{{\bf{Q}}_c},{{\bf{Q}}_a}) - {\phi _k}({{\bf{Q}}_c},{{\bf{Q}}_a}) \ge {\tau_{ms}}, \forall k \in {\cal{K}}\label{op4b}\\
&\text{Tr}({{\bf{Q}}_0} + {{\bf{Q}}_c}+{{\bf{Q}}_a}) \le P, \\
&{{\bf{Q}}_0} \succeq {\bf{0}}, {{\bf{Q}}_a} \succeq {\bf{0}}, {{\bf{Q}}_c} \succeq {\bf{0}}.
\end{align}
\end{subequations}

Since ${\phi _k}({{\bf{Q}}_c},{{\bf{Q}}_a})$, ${\varphi _k}({{\bf{Q}}_a})$ and ${\eta _k}({{\bf{Q}}_0},{{\bf{Q}}_c},{{\bf{Q}}_a})$ are all concave w.r.t. $({{\bf{Q}}_0},{{\bf{Q}}_c},{{\bf{Q}}_a})$, one can easily notice that the objective function of (\ref{op2}) and constraints (\ref{op4a}) and (\ref{op4b}) are all in a difference-of-concave form. This property makes problem (\ref{op1}) fall into the context of DC program \cite{NIPS2009}, which can be iteratively solved via DC algorithm.

Our next endeavor is to show the DC approach to (\ref{op4}) mathematically. Its basic idea is to locally linearize the nonconcave parts in (\ref{op4}) at some feasible point via Taylor series expansion (TSE), and then iteratively solve the linearized problem. To this end, we introduce the TSE via the following lemma.
\begin{lemma}[\cite{chu2015secrecy}]\label{TSE}
An affine Taylor series approximation of a function $f({\mathbf{X}}):{{\mathbb R}^{M \times N}} \to {\mathbb R}$ can be expressed at ${\bf{\tilde X}}$ as below.
\begin{equation}
f\left( {\bf{X}} \right) \approx f( {{\bf{\tilde X}}}) + {\rm{vec}}\left( {f'\left( {\bf{X}} \right)} \right)^H{\rm{vec}}({{\bf{X}} - {\bf{\tilde X}}} ).
\end{equation}
\end{lemma}

The TSE above enables us to reformulate the primal nonconcave parts of (\ref{op4}) into a linear form. In particular, by applying Lemma \ref{TSE} and the fact $\partial \left( {\log \left| {\bf{X}} \right|} \right) = {\rm{Tr}}\left( {{{\bf{X}}^{ - 1}}\partial {\bf{X}}} \right)$, ${\varphi _1}({{\bf{Q}}_a})$ can be approximated as
\begin{align}
\nonumber{\varphi _1}({{\bf{Q}}_a})&=\log \left| {{\bf{I}} + {{\bf{H}}_1}{{\bf{Q}}_a}{\bf{H}}_1^H} \right|\\
\nonumber&\approx {\varphi _1}({{\bf{\tilde Q}}_a})+ ({\rm{vec}}\left({\bf{S}}\right))^H{\rm{vec}}\left( {{{\bf{Q}}_a} - {{{\bf{\tilde Q}}}_a}} \right)\\
\nonumber&\mathop= \limits^{(a)} {\varphi _1}({{\bf{\tilde Q}}_a})+ {\rm{Tr}}\left[{\bf{S}}({{\bf{Q}}_a}-{{{\bf{\tilde Q}}}_a})\right],\\
&\buildrel \Delta \over = {\tilde \varphi _1}({{\bf{Q}}_a}) \label{approx1}
\end{align}
in the objective function of (\ref{op4}), where ${{{\bf{\tilde Q}}}_a}$ is a given transmit covariance matrix, ${\bf{S}} \buildrel \Delta \over = {{{\bf{H}}_1^H}{\left( {{\bf{I}} + {{\bf{H}}_1}{{{\bf{\tilde Q}}}_a}{\bf{H}}_1^H} \right)^{ - 1}}{\bf{H}}_1}$ and the equality $(a)$ is due to the fact that ${\rm{Tr}}({{\bf{A}}^H}{\bf{B}}) = {({\rm{vec}}({\bf{A}}))^H}{\rm{vec}}({\bf{B}})$ for appropriate dimensions of ${\bf{A}}$ and ${\bf{B}}$. Likewise, ${\phi _k}({{\bf{Q}}_c},{{\bf{Q}}_a})$, appearing in the constraints (\ref{op4a}) and (\ref{op4b}), can be approximated as
\begin{align}
&\nonumber{\phi _k}({{\bf{Q}}_c},{{\bf{Q}}_a})=\log \left| {{\bf{I}} + {{\bf{H}}_k}({{\bf{Q}}_c} + {{\bf{Q}}_a}){\bf{H}}_k^H} \right| \\
\nonumber&\approx {\phi _k}({{\bf{\tilde Q}}_c},{{\bf{\tilde Q}}_a})+{\rm{Tr}}\left[{\bf{U}}({{\bf{Q}}_c}-{{{\bf{\tilde Q}}}_c})\right]+{\rm{Tr}}\left[{\bf{U}}({{\bf{Q}}_a}-{{{\bf{\tilde Q}}}_a})\right]\\
&\buildrel \Delta \over = {\tilde \phi _k}({{\bf{Q}}_c},{{\bf{Q}}_a}),\label{approx2}
\end{align}
in which ${\bf{U}}$ is determined by
\begin{equation}
{\bf{U}} = {\bf{H}}_k^H{({\bf{I}} + {{\bf{H}}_k}({{{\bf{\tilde Q}}}_c} + {{{\bf{\tilde Q}}}_a}){\bf{H}}_k^H)^{ - 1}}{{\bf{H}}_k}.
\end{equation}

Based on the approximations above, the original QoMS-constrained SRM problem (\ref{op4}) can be reformulated as
\begin{subequations}\label{op5}
\begin{align}
\nonumber &{\bar{R}}(\tau _{ms}) =\mathop{\max}\limits_{{{\bf{Q}}_0},{{\bf{Q}}_c},{{\bf{Q}}_a},\eta} {\phi _1}({{\bf{Q}}_c},{{\bf{Q}}_a}) - {\tilde \varphi _1}({{\bf{Q}}_a})-\eta\\
\text{s.t.}\; &{\varphi _k}({{\bf{Q}}_a})-{\tilde \phi _k}({{\bf{Q}}_c},{{\bf{Q}}_a})+ \eta \ge 0, \forall k \in {\cal{K}}_e\label{op5a}\\
&{\eta _k}({{\bf{Q}}_0},{{\bf{Q}}_c},{{\bf{Q}}_a}) - {\tilde \phi _k}({{\bf{Q}}_c},{{\bf{Q}}_a}) \ge {\tau_{ms}}, \forall k \in {\cal{K}}\label{op5b}\\
&\text{Tr}({{\bf{Q}}_0} + {{\bf{Q}}_c}+{{\bf{Q}}_a}) \le P, \\
&{{\bf{Q}}_0} \succeq {\bf{0}}, {{\bf{Q}}_a} \succeq {\bf{0}}, {{\bf{Q}}_c} \succeq {\bf{0}}.
\end{align}
\end{subequations}
where ${\bar{R}}(\tau _{ms})$ is the optimal objective value of (\ref{op3}), serving as an approximation to $R(\tau _{ms})$. According to the relationship between a concave function and its Taylor series expansion, it is immediate to get
\begin{align}
&{\varphi _1}({{\bf{Q}}_a}) \le {\tilde \varphi _1}({{\bf{Q}}_a}), \forall {{\bf{Q}}_a} \succeq {\bf{0}},\nonumber\\
&{\phi _k}({{\bf{Q}}_c},{{\bf{Q}}_a}) \le {\tilde \phi _k}({{\bf{Q}}_c},{{\bf{Q}}_a}), \forall {{\bf{Q}}_a} \succeq {\bf{0}}, {{\bf{Q}}_c} \succeq {\bf{0}}.\label{relation}
\end{align}
As a consequence, any feasible solution to (\ref{op5}) should also be feasible to (\ref{op4}), and ${\bar{R}}(\tau _{ms}) \le {R}(\tau _{ms})$ must hold.

This approximated problem (\ref{op5}) is convex with regard to (w.r.t.) $\left({{\bf{Q}}_0},{{{\bf{Q}}_c},{{\bf{Q}}_a}} \right)$ and hence $\left({{\bf{Q}}_0},{{{\bf{Q}}_c},{{\bf{Q}}_a}} \right)$ can be iteratively obtained by solving problem (\ref{op5}) via some off-the-shelf interior-point algorithm, e.g., \texttt{CVX}. We summarize our proposed iterative algorithm for solving (\ref{op4}) in Algorithm 1. To acquire the secrecy rate region, we need to traverse ${\tau_{ms}}$ lying within the interval $[0,\tau_{\max}]$ and store the corresponding objective value of (\ref{op5}).
\begin{algorithm}
  \caption{Iterative method for solving (\ref{op4})}
  \begin{algorithmic}[1]\label{DC.Alg}
    \State Initiate $n=0$ and choose an arbitrary starting point $({{\bf{\tilde Q}}_{c,n}},{{\bf{\tilde Q}}_{a,n}})$ feasible to (\ref{op5})
    \State \textbf{Repeat}
    \State \quad Solve (\ref{op5}) with ${\bf{\tilde Q}}_c={{\bf{\tilde Q}}_{c,n}}$ and ${\bf{\tilde Q}}_a={{\bf{\tilde Q}}_{a,n}}$, and obtain $({{\bf{Q}}_c^*},{{\bf{Q}}_a^*})$, which is the optimal solution of (\ref{op5});\label{sub}
    \State \quad Update ${{\bf{\tilde Q}}_{c,n+1}}={{\bf{Q}}_c^*}$, ${{\bf{\tilde Q}}_{a,n+1}}={{\bf{Q}}_a^*}$;
    \State \quad Update $n=n+1$;
    \State \textbf{Until the convergence conditions are satisfied.}
    \State Output ${{\bf{\tilde Q}}_{c,n}}$ and ${{\bf{\tilde Q}}_{a,n}}$.
  \end{algorithmic}
\end{algorithm}

\begin{remark}\label{initial}
In Algorithm 1, the initialization of $({{\bf{\tilde Q}}_{c,0}},{{\bf{\tilde Q}}_{a,0}})$ plays a crucial role in influencing the total iteration times. Let us define $\left({{\bf{Q}}^{i}_c},{{\bf{Q}}^{i}_a}\right)$ as the output solution in $i$th traversal of ${\tau_{ms}}$. The following ``warmstart operation'' could be adopted to initialize $({{\bf{\tilde Q}}_{c,0}},{{\bf{\tilde Q}}_{a,0}})$ for achieving a fast convergence rate:

\textbf{Warmstart Operation}: We start the traversal of ${\tau_{ms}}$ from ${\tau_{ms}}={\tau_{\max}}$. In the first traversal, ${{\bf{\tilde Q}}_{c,0}}$ and ${{\bf{\tilde Q}}_{a,0}}$ are both initialized as $\bf{0}$. In the $i$th ($i>1$) traversal, $({{\bf{\tilde Q}}_{c,0}},{{\bf{\tilde Q}}_{a,0}})$ is initialized as the solution output by Algorithm 1 in the $(i-1)$th traversal.
\end{remark}

\subsection{Convergence Analysis}
As one can see, the basic merit of DC lies in its tractability, which caters to the numerical optimization using the parser-solver. As an additional merit, the proposed DC approach has a theoretically provable guarantee on its solution convergence, which will be demonstrated in the following proposition.
\begin{proposition}
Every limit point of $\left( {{{\bf{Q}}_0^*},{{\bf{Q}}_c^*}} \right)$ is a stationary point of problem (\ref{op2})
\end{proposition}
\begin{IEEEproof}
The proof is a direct application of \cite[Th 10]{NIPS2009}, and thus omitted here for simplicity.
\end{IEEEproof}

\section{An AO-Based Approach to the SRRM Problem}
In this section, we develop our another scalarization method, referred to as weighted-sum method, to solve (\ref{op1}). The basic problem formulation is a WSRM problem, which can be solved via an AO-based approach. Here we should point out that the application of AO to SRM problem has been observed in some existing papers, i.e., \cite{li2013transmit}. Nonetheless, the AO algorithm we used in this section is a nontrivial extension of that in \cite{li2013transmit}. Specifically, the objective function in \cite{li2013transmit} only contains a single secrecy rate term. While in our considered scenario, an extra multicast rate term is incorporated, which brings some new issues, say, the convergence proof, that should be tackled.

\subsection{Scalarization}
The basic idea of the weighted-sum method is to introduce a so-called weight vector \cite{boyd2009convex} that is positive in the dual cone $K^*={\mathbb {R}}_ + ^2$, and then to transform the primal vector optimization problem into a scalar optimization problem. By varying the vector, we can obtain different Pareto optimal solutions of (\ref{op1}).

To put into context, the Pareto boundary of (\ref{Region1}) can be characterized by the solution of
\begin{equation}\label{op6}
\begin{split}
&\mathop{\max}\limits_{{{\bf{Q}}_0},{{\bf{Q}}_c},{{\bf{Q}}_a},R_0,R_c} R_0 + \lambda_c R_c\\
\text{s.t.}\quad &{R_0} \le \mathop {\min }\limits_{k \in {\cal K}} C_{m,k}({{\bf{Q}}_0},{{\bf{Q}}_c},{{\bf{Q}}_a})\\
&{R_c} \le  C_b({{\bf{Q}}_c},{{\bf{Q}}_a}) - \mathop {\max}\limits_{k \in {{\cal K}_e}} C_{e,k}({{\bf{Q}}_c},{{\bf{Q}}_a})\\
&\text{(\ref{op1a})-(\ref{op1b}) satisfied},
\end{split}
\end{equation}
in which $\lambda_c \in [0,+\infty)$ and ${\bm{\lambda }} = [1,\lambda_c]$ is our introduced weight vector. In general, the optimal $\left(R_0,R_c\right)$ to (\ref{op2}) is the point where a straight line with slope $-1/{\lambda_c}$ is tangent to the Pareto boundary. Before proceeding, let us first point out some special cases of problem (\ref{op2}).
\begin{enumerate}
  \item When ${\bm{\lambda }} = [1,1]$, the optimal $\left(R_0,R_c\right)$ turns out to be the so-called utilitarian point, also referred to as ``sum-rate'' point in communications.
  \item The single-service points are the two points where $R_0 = 0$ and where $R_c = 0$, respectively. When $R_0 = 0$, problem (\ref{op2}) is degraded into a conventional AN-aided SRM problem in MIMO wiretap channel. When $R_c = 0$, the maximum $R_0$ can be derived by solving the same convex optimization problem as (\ref{max_tau}).
\end{enumerate}

\subsection{AO Iterative Algorithm}
We are now in a position to determine the tractable approaches to the WSRM problem (\ref{op6}). First, one can notice that by discarding $R_0$ and $R_c$ as slack variables, problem (\ref{op6}) is equivalent to the following optimization problem.
\begin{subequations}\label{op7}
\begin{align}
\nonumber R(\lambda_c)=&\mathop{\max}\limits_{{{\bf{Q}}_0},{{\bf{Q}}_c},{{\bf{Q}}_a}} \lambda_c (C_b - \mathop {\max}\limits_{k \in {{\cal K}_e}} C_{e,k}) + \mathop {\min }\limits_{k \in {\cal K}} C_{m,k}\\
\text{s.t.}\quad &\text{Tr}({{\bf{Q}}_0} + {{\bf{Q}}_c} + {\bf{Q}}_a) \le P,\label{op7a}\\
&{{\bf{Q}}_0} \succeq {\bf{0}}, {{\bf{Q}}_c} \succeq {\bf{0}}, {\bf{Q}}_a \succeq {\bf{0}}.\label{op7b}
\end{align}
\end{subequations}

The obstacle of solving (\ref{op7}) mainly lies in the non-smoothness of its objective function, which negates the use of many derivative-related iterative algorithms. As a result, we next develop a derivative-free AO iterative algorithm to solve (\ref{op7}). To this end, we will first need to transform the WSRM problem (\ref{op7}) into a form amenable to AO.

\begin{lemma}[\cite{li2013transmit}]\label{lem1}
Let ${\bf{E}} \in {{\mathbb{C}}^{N \times N}}$ be any matrix satisfying ${\bf{E}} \succ \mathbf{0}$. Define the function $f({\bf{S}}) =  - {\rm{Tr}}({\bf{SE}}) + \log \left| {\bf{S}} \right| + N$. Then
\begin{equation}\label{lemma1}
\log \left| {{{\bf{E}}^{ - 1}}} \right| = \mathop {\max }\limits_{{\bf{S}} \in {{\mathbb{C}}^{N \times N}},{\bf{S}} \succeq 0} f({\bf{S}}),
\end{equation}
and the optimal solution to the right-hand side (RHS) of (\ref{lemma1}) is ${{\bf{S}}^ * } = {{\bf{E}}^{ - 1}}$.
\end{lemma}

Applying Lemma \ref{lem1} to ${C_b}$, $C_{e,k}$ and $C_{m,k}$, one can obtain
\begin{subequations}\label{eq1}
\begin{align}
{C_b}&=\mathop {\max }\limits_{{{\bf{S}}_1} \succeq {\bf{0}}} {\varphi _b}({{\bf{Q}}_c},{{\bf{Q}}_a},{\bf{S}}_1),\label{eq1a}\\
C_{e,k}&=\mathop {\min }\limits_{{{\bf{S}}_k} \succeq {\bf{0}}} {\varphi _{e,k}}({{\bf{Q}}_c},{{\bf{Q}}_a},{\bf{S}}_k), \forall k \in {\cal{K}}_e,\label{eq1b}\\
C_{m,k}&=\mathop {\max }\limits_{{{\bf{U}}_k} \succeq {\bf{0}}} {\varphi _{m,k}}({{\bf{Q}}_0}, {{\bf{Q}}_c},{{\bf{Q}}_a},{\bf{U}}_k), \forall k \in {\cal{K}},\label{eq1c}
\end{align}
\end{subequations}
where we define
\begin{subequations}\label{eq2}
\begin{align}
\nonumber&{\varphi _b}({{\bf{Q}}_c},{{\bf{Q}}_a},{\bf{S}}_1)= - {\rm{Tr}}({{\bf{S}}_1}({\bf{I}} + {{\bf{H}}_1}{{\bf{Q}}_a}{\bf{H}}_1^H)+ \log\left| {{{\bf{S}}_1}} \right|+N_{r,1} \\
&+ \log \left| {{\bf{I}} + {{\bf{H}}_1}({{\bf{Q}}_a} + {{\bf{Q}}_c}){\bf{H}}_1^H} \right|,\\
\nonumber&{\varphi _{e,k}}({{\bf{Q}}_c},{{\bf{Q}}_a},{\bf{S}}_k)= - \log \left| {{{\bf{S}}_k}} \right| - \log \left| {{\bf{I}}+{{\bf{H}}_k}{{\bf{Q}}_a}{\bf{H}}_{k}^H} \right|-N_{r,k} \\
&+{\rm{Tr}}({{\bf{S}}_k}({\bf{I}} + {{\bf{H}}_{k}}({{\bf{Q}}_a} + {{\bf{Q}}_c}){\bf{H}}_{k}^H)),\\
\nonumber&{\varphi _{m,k}}({{\bf{Q}}_0}, {{\bf{Q}}_c},{{\bf{Q}}_a},{\bf{U}}_k)=- {\rm{Tr}}({{\bf{U}}_k}({\bf{I}} + {{\bf{H}}_k}({{\bf{Q}}_c}+{{\bf{Q}}_a}){\bf{H}}_k^H) \\
&+ \log\left| {{{\bf{U}}_k}} \right|+ \log \left| {{\bf{I}} + {{\bf{H}}_k}({{\bf{Q}}_0} + {{\bf{Q}}_c} + {{\bf{Q}}_a}){\bf{H}}_k^H} \right|+ N_{r,k},
\end{align}
\end{subequations}
in which $\{{\bf{S}}_k\}_{k \in {\cal{K}}}$ and $\{{\bf{U}}_k\}_{k \in {\cal{K}}}$ are slack variables satisfying ${\bf{S}}_k \succeq \mathbf{0}$ and ${\bf{U}}_k \succeq \mathbf{0}$ for $\forall k \in {\cal{K}}$.

Following the matrix manipulations in \cite{li2013transmit}, we have
\begin{equation}\label{eq3}
\begin{split}
&\mathop {\max }\limits_{k \in {\cal{K}}_e} \mathop {\min }\limits_{{{\bf{S}}_k}} {\varphi _{e,k}}({{\bf{Q}}_c},{{\bf{Q}}_a},{\bf{S}}_k)\\
=&\mathop {\min }\limits_{\{ {{\bf{S}}_k}\} _{k \in {\cal{K}}_e}} \mathop {\max }\limits_{k \in {\cal{K}}_e} {\varphi _{e,k}}({{\bf{Q}}_c},{{\bf{Q}}_a},{\bf{S}}_k),
\end{split}
\end{equation}
and
\begin{equation}\label{eq4}
\begin{split}
&\mathop {\min }\limits_{k \in {\cal{K}}} \mathop {\max }\limits_{{{\bf{U}}_k}} {\varphi _{m,k}}({{\bf{Q}}_0},{{\bf{Q}}_c},{{\bf{Q}}_a},{\bf{U}}_k)\\
=& \mathop {\max }\limits_{\{ {{\bf{U}}_k}\} _{k \in {\cal{K}}}} \mathop {\min }\limits_{k \in {\cal{K}}} {\varphi _{m,k}}({{\bf{Q}}_0},{{\bf{Q}}_c},{{\bf{Q}}_a},{\bf{U}}_k).
\end{split}
\end{equation}
Substituting (\ref{eq1a})-(\ref{eq1c}) into (\ref{op7}) and making use of (\ref{eq3}) and (\ref{eq4}), one can check that problem (\ref{op7}) is equivalent to the following optimization problem.
\begin{subequations}\label{op8}
\begin{align}
\nonumber R(\lambda_c)=&\mathop {\max }\limits_{{{\bf{Q}}_0},{{\bf{Q}}_c},{{\bf{Q}}_a},\atop \left\{ {\bf{S}}_k \right\}_{k \in {\cal{K}}}, \left\{ {\bf{U}}_k \right\}_{k \in {\cal{K}}}}f({{\bf{Q}}_0},{{\bf{Q}}_c},{{\bf{Q}}_a},\left\{ {\bf{S}}_k \right\}_{k \in {\cal{K}}},\left\{ {\bf{U}}_k \right\}_{k \in {\cal{K}}})\\
\text{s.t.}\quad&\text{Tr}({{\bf{Q}}_0} + {{\bf{Q}}_c} + {\bf{Q}}_a) \le P,\label{op8a}\\
&{{\bf{Q}}_0} \succeq {\bf{0}}, {{\bf{Q}}_c} \succeq {\bf{0}}, {\bf{Q}}_a \succeq {\bf{0}},\label{op8b}
\end{align}
\end{subequations}
in which we define
\begin{equation}\label{eq5}
\begin{split}
&f({{\bf{Q}}_0},{{\bf{Q}}_c},{{\bf{Q}}_a},\left\{ {\bf{S}}_k \right\}_{k \in {\cal{K}}},\left\{ {\bf{U}}_k \right\}_{k \in {\cal{K}}}) = \\ &\lambda_c ({\varphi _b}({{\bf{Q}}_c},{{\bf{Q}}_a},{\bf{S}}_1)- \mathop {\max }\limits_{k \in {\cal{K}}_e}{\varphi _{e,k}}({{\bf{Q}}_c},{{\bf{Q}}_a},\{ {{\bf{S}}_k}\} _{k \in {\cal{K}}_e}))\\
&+ \mathop {\min }\limits_{k \in {\cal{K}}}{\varphi _{m,k}}({{\bf{Q}}_0},{{\bf{Q}}_c},{{\bf{Q}}_a},\left\{ {\bf{U}}_k \right\}_{k \in {\cal{K}}}).
\end{split}
\end{equation}

The upshot of this reformation is that problem (\ref{op7}) becomes primal decomposable. Specifically, problem (\ref{op8}) is convex w.r.t. \emph{either} $({{\bf{Q}}_0},{{\bf{Q}}_c},{{\bf{Q}}_a})$ \emph{or} $(\left\{ {\bf{S}}_k \right\}_{k \in {\cal{K}}},\left\{ {\bf{U}}_k \right\}_{k \in {\cal{K}}})$. Hence, AO is naturally employed to solve (\ref{op8}). With $({{\bf{Q}}_0},{{\bf{Q}}_c},{{\bf{Q}}_a})$ fixed, the optimal solution of $(\left\{ {\bf{S}}_k \right\}_{k \in {\cal{K}}},\left\{ {\bf{U}}_k \right\}_{k \in {\cal{K}}})$ admits an analytical expression, according to Lemma \ref{lem1}, given by
\begin{subequations}\label{cls}
\begin{align}
&{\bf{S}}_1^ *  = {({\bf{I}} + {{\bf{H}}_1}{{\bf{Q}}_a}{\bf{H}}_1^H)^{ - 1}},\\
&{\bf{S}}_k^ *  = {({\bf{I}} + {{\bf{H}}_k}({{\bf{Q}}_a} + {{\bf{Q}}_c}){\bf{H}}_k^H)^{ - 1}}, \forall k \in {\cal{K}}_e,\\
&{\bf{U}}_k^ *  = {({\bf{I}} + {{\bf{H}}_k}({{\bf{Q}}_a} + {{\bf{Q}}_c}){\bf{H}}_k^H)^{ - 1}}, \forall k \in {\cal{K}},
\end{align}
\end{subequations}
in which we utilize the fact that $\left\{ {\bf{S}}_k \right\}_{k \in {\cal{K}}}$ and $\left\{ {\bf{U}}_k \right\}_{k \in {\cal{K}}}$ are decoupled among ${\varphi _b}$, ${\varphi _{e,k}}$ and ${\varphi _{m,k}}$. Comparatively, with $(\left\{ {\bf{S}}_k \right\}_{k \in {\cal{K}}},\left\{ {\bf{U}}_k \right\}_{k \in {\cal{K}}})$ fixed, the optimal solution of $({{\bf{Q}}_0},{{\bf{Q}}_c},{{\bf{Q}}_a})$ can be obtained by solving a convex optimization problem as below, i.e.,
\begin{equation}\label{op9}
\begin{split}
({\bf{Q}}_0^ * &,{\bf{Q}}_c^ * ,{\bf{Q}}_a^ *) = \\
&\arg \mathop {\max }\limits_{({{\bf{Q}}_0},{{\bf{Q}}_c},{{\bf{Q}}_a}) \in {\cal{F}}} f({{\bf{Q}}_0},{{\bf{Q}}_c},{{\bf{Q}}_a},\left\{ {\bf{S}}_k \right\}_{k \in {\cal{K}}},\left\{ {\bf{U}}_k \right\}_{k \in {\cal{K}}}),
\end{split}
\end{equation}
where ${\cal{F}}$ denotes the feasible set of (\ref{op7}), which is convex.

The whole AO process for solving (\ref{op8}) is given in Algorithm 2. In line \ref{cp} of Algorithm 2, the convex subproblem can be solved via \texttt{CVX}. Following the similar warmstart operation introduced in Remark \ref{initial}, the iteration times of Algorithm 2 can be significantly decreased.
\begin{algorithm}
  \caption{AO algorithm for solving (\ref{op8})}
  \begin{algorithmic}[1]\label{AO.Alg}
    \State Initiate $n=1$, and $({\bf{Q}}_c^0,{\bf{Q}}_a^0) \in {\cal{F}}$;
    \State \textbf{Repeat}
    \State \quad ${\bf{S}}_1^ {n}  = {({\bf{I}} + {{\bf{H}}_1}{{\bf{Q}}_a^{n-1}}{\bf{H}}_1^H)^{ - 1}}$;
    \State \quad ${\bf{S}}_k^ {n} = ({\bf{I}} + {{\bf{H}}_k}({{\bf{Q}}_a^{n-1}} + {{\bf{Q}}_c^{n-1}}){\bf{H}}_k^H)^{ - 1}, \forall k \in {\cal{K}}_e$;
    \State \quad ${\bf{U}}_k^ {n} = ({\bf{I}} + {{\bf{H}}_k}({{\bf{Q}}_a^{n-1}} + {{\bf{Q}}_c^{n-1}}){\bf{H}}_k^H)^{ - 1}, \forall k \in {\cal{K}}$;
    \State \quad $({\bf{Q}}_0^ {n},{\bf{Q}}_c^ {n},{\bf{Q}}_a^ {n}) = \arg \mathop {\max }\limits_{({{\bf{Q}}_0},{{\bf{Q}}_c},{{\bf{Q}}_a}) \in {\cal{F}}} f({{\bf{Q}}_0},{{\bf{Q}}_c},{{\bf{Q}}_a},$ \label{cp}
    $\left\{ {{{\bf{S}}_k^{n}}} \right\}_{k \in {\cal{K}}}, \left\{ {{{\bf{U}}_k^{n}}} \right\}_{k \in {\cal{K}}})$;
    \State \quad $n=n+1$;
    \State \textbf{Until the convergence conditions are satisfied.}
    \State Output $({\bf{Q}}_0^n,{\bf{Q}}_c^n,{\bf{Q}}_a^n)$.
  \end{algorithmic}
\end{algorithm}

\subsection{Convergence Analysis}
It can be verified that the AO algorithm produces a nondecreasing objective value of (\ref{op8}). Besides, the following convergence result is always guaranteed.
\begin{proposition}\label{KKTConv}
Suppose that $({\bf{Q}}_0^ n ,{\bf{Q}}_c^ n ,{\bf{Q}}_a^ n)$ is the solution generated by the AO algorithm in $n^{\text{th}}$ iteration, then the sequence $\{({\bf{Q}}_0^ n ,{\bf{Q}}_c^ n ,{\bf{Q}}_a^ n)\}_n$ must converge to one stationary point (i.e., Karush-Kuhn-Tucker (KKT) point) of the primal WSRM problem (\ref{op7}).
\end{proposition}
\begin{IEEEproof}
The proof can be found in Appendix \ref{AO_Appendix}.
\end{IEEEproof}

Since the global optimal solutions to problem (\ref{op7}) hitherto remains inaccessible, our achieved secrecy rate region would serve as a lower bound on $R_s({\left\{ {{\bf{H}}_k} \right\}_{k \in {\cal{K}}}},P)$, which achieves KKT optimality.

\section{Comparison of the Proposed Methods}
In the previous sections, we present two tractable convex formulations of the SRRM problem (\ref{op1}). This naturally leads to the question about the relative performance of the two formulations. In the following subsections, we address this question by comparing their performance and computational complexity in solving (\ref{op1}).

\subsection{Performance Analysis}
As introduced in the preceding sections, the QoMS-based scalarization can yield a complete set of boundary points of $R_s({\left\{ {{\bf{H}}_k} \right\}_{k \in {\cal{K}}}},P)$, which contains all Pareto optimal points of (\ref{op1}). The resulting scalar problem (\ref{relax}) aims to maximize the secrecy rate and meanwhile maintain the QoMS above a given threshold. Predictably, the use of AN should be effective merely at low QoMS region, since AN exerts a negative effect on the multicasting performance. To guarantee the high demand for QoMS, AN has to be prohibitive at high QoMS region. This QoMS-constrained SRM is a generalization of traditional SRM in physical-layer security, and provides the transmitter with some insights in how to tradeoff the security performance and the multicasting performance.

As for the weighted-sum scalarization method, the necessary condition for it to find all Pareto optimal points is that the secrecy rate region should be convex. Besides, its performance is also dependent on the precision of $\lambda_c$. The traversal of $\lambda_c$ should span from zero to an extremely large number with appropriate step, so that each Pareto optimal points can be detected. Nonetheless, the weighted-sum problem structure has an interesting pricing interpretation from the field of economics. To elaborate a little further, let us define $p_0$ and $p_c$ as the unit price for the secrecy rate and the multicast rate, respectively, charged by the service provider. To maximize its revenue, the service provider should be concerned about how to solve the WSRM problem in (\ref{op6}) with setting $\lambda_c=p_c/p_0$. The use of AN could also be explained in this context. It is evident to see when $p_0 \gg p_c$, the revenue from multicasting transmission would dominate the objective function of (\ref{op6}), and thus, eliminating AN would be helpful in increasing the overall revenue.

In all, these two scalarization methods are suitable for different application scenarios and provide different insights. Nonetheless, the QoMS-based scalarization could yield all Pareto optimal points, while the weighted-sum scalarization might only yield some of them, dependent on the shape of the secrecy rate region.

\begin{remark}
Besides the QoMS-based and weighted-sum scalarization methods, some other scalarization methods have been proposed in literature to find the complete Pareto set for biobjective optimization, e.g., the weighted Tchebycheff method \cite{marler2004survey}. However, to implement this method, one has to first obtain the single-service point of the confidential message (cf. (\ref{single1})) and then solve a highly nonconvex max-min optimization problem.
\begin{align}
{R_c^{\max}} =& \mathop {\max }\limits_{{{\bf{Q}}_c} \succeq {\bf{0}},{\rm{Tr}}({{\bf{Q}}_c}) \le P} \log \left| {{\bf{I}} + {{\bf{H}}_1}{{\bf{Q}}_c}{\bf{H}}_1^H} \right| \nonumber\\
&- \mathop {\max }\limits_{k \in {\cal K}} \log \left| {{\bf{I}} + {{\bf{H}}_k}{{\bf{Q}}_c}{\bf{H}}_k^H} \right|.\label{single1}
\end{align}
Unfortunately, problem (\ref{single1}) is nonconvex, and so the optimal solution to (\ref{single1}) may not be obtained, which invalidates the use of the weighted Tchebycheff method.
\end{remark}

\subsection{Complexity Analysis}
The major computational complexity of the two scalarization methods comes from solving the problems (\ref{op5}) and (\ref{op9}). While
both of problems (\ref{op5}) and (\ref{op9}) are convex, they are not in a standard semidefinite programming (SDP) form, owing to the logarithm functions therein. To solve them, a successive approximation method embedded with a primal-dual interior-point method (IPM) is employed, say by \texttt{CVX}. As is known, the arithmetic complexity for the generic primal-dual IPM to solve a standard SDP is ${\cal O}(\max {\{ m,n\} ^4}{n^{1/2}}\log (1/\varepsilon ))$\cite{luo2010semi}, in which $m$, $n$ and $\varepsilon$ represent the number of linear constraints, the dimension of the positive semidefinite cone and the solution accuracy, respectively. Therefore, the complexity of solving (\ref{op5}) or (\ref{op9}) is ${\cal O}({L_{SA}}\max {\{ 2K,{N_t}\} ^4}N_t^{1/2}\log (1/\varepsilon ))$, where $L_{SA}$ denotes the number of successive approximations used. Since we are not aware of the relation between $L_{SA}$ and $N_t$, this complexity expression is rather rough.

However, by utilizing the following approximation \cite{cumanan2014secrecy}:
\begin{equation}\label{approx3}
\log \left| {{\bf{I}} + {\bf{HQ}}{{\bf{H}}^H}} \right| = {\rm{Tr}}({\bf{HQ}}{{\bf{H}}^H}) + {\cal O}(\left\| {{\bf{HQ}}{{\bf{H}}^H}} \right\|),
\end{equation}
all logarithm terms in problems (\ref{op5}) and (\ref{op9}) can be approximated by a trace function at \emph{low} transmit power. This approximation further converts the convex problems (\ref{op5}) and (\ref{op9}) into SDP ones, which makes it possible to acquire a more accurate big-O expression of the computational complexity for low transmit power.

Specifically, consider (\ref{op5}), which has three linear matrix inequality (LMI) constraints of size $N_t$, and $2K$ LMI constraints of size 1 after introducing the approximation (\ref{approx3}). Moreover, for (\ref{op5}), the number of decision variables is on the order $n_1=3N_t^2+1$. Then, when a generic path-following IPM is used to solve problem (\ref{op5}), the total arithmetic computation cost is on the order of \cite{ben2001lectures}
\begin{equation}\label{order1}
\begin{split}
&{T_1} = \sqrt {2K + 3{N_t}}\phi(n_1),\\
&\phi(n_1)={{n_1}(2K + 3N_t^3) + n_1^2(2K + 3N_t^2) + n_1^3}
\end{split}
\end{equation}
with $n_1={\cal O}(3N_t^2+1)$.

On the other hand, for solving (\ref{op9}), we need to introduce two additional slack variables to move the maximum and minimum terms in the objective function of (\ref{op9}) to the constraints. Hence, the number of decision variables is on the order of $n_2=3N_t^2+2$, and (\ref{op9}) also has three LMI constraints of size $N_t$, and $2K$ LMI constraints of size 1. The total arithmetic computation cost for solving (\ref{op9}) is on the order of
\begin{equation}\label{order2}
\begin{split}
&{T_2} = \sqrt {2K + 3{N_t}}\phi(n_2),\\
&\phi(n_2)={{n_2}(2K + 3N_t^3) + n_2^2(2K + 3N_t^2) + n_2^3}
\end{split}
\end{equation}
with $n_2={\cal O}(3N_t^2+2)$.

Comparing (\ref{order1}) and (\ref{order2}), one can note that the total arithmetic computation cost of solving the two problems is comparable, with $T_2$ slightly greater than $T_1$ due to $n_2 > n_1$. This observation implies that the QoMS-based scalarization is more time-efficient at low transmit power. This is also consistent with our following simulation results, as we shall see in Section \Rmnum{6}.

\section{Numerical Results}
In this section, we provide numerical results to illustrate the secrecy rate region derived from the two proposed methods, compared with two other existing strategies. The first one is the no-AN transmission, i.e., prefixing ${{\bf{Q}}_a}$ as $\bf{0}$ in problem (\ref{op1}). Thus, its achieved secrecy rate region can also be derived via the DC and AO algorithms. Another one is the traditional service integration using time division multiple address (TDMA), which assigns the confidential message and multicast message to two orthogonal time slots. Its maximum secrecy rate and multicast rate can be obtained by seeking the single-service points of $R_s({\left\{ {{\bf{H}}_k} \right\}_{k \in {\cal{K}}}},P)$. For the fairness of comparison, the secrecy rate and multicast rate achieved by this TDMA-based strategy should be \textbf{halved}\cite{Wyrembelski2012Physical}.

In the first subsection, the convergence results of both algorithms are presented. The second subsection gives the comparison between these two algorithms in terms of achievable performance and computational complexity.

\subsection{Convergence Results}
In this subsection, we assume $N_t=5$, $N_{r,k}=3$ for all $k \in \cal{K}$, and $K=4$. The channel matrices are randomly generated from an i.i.d. complex Gaussian distribution with zero mean and unit variance. According to Proposition \ref{P1}, since $N_t > N_{r,1}$, the optimal solution to (\ref{relax}) is attained when the constraint (\ref{relax.a}) holds with equality.

\begin{figure}[!t]
\begin{center}
\includegraphics[width=3in]{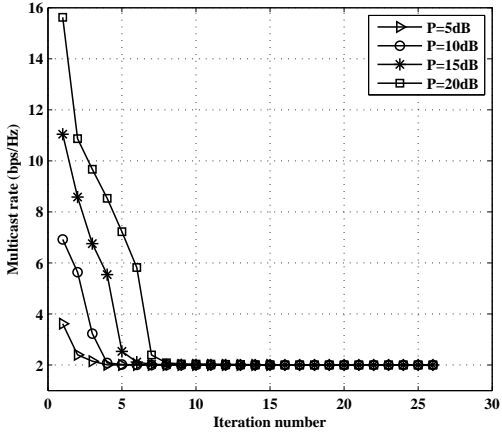}
\caption{DC algorithm: Convergence of the multicast rate}\label{Convergence_Multi}
\end{center}
\end{figure}
\begin{figure}[!t]
\begin{center}
\includegraphics[width=3in]{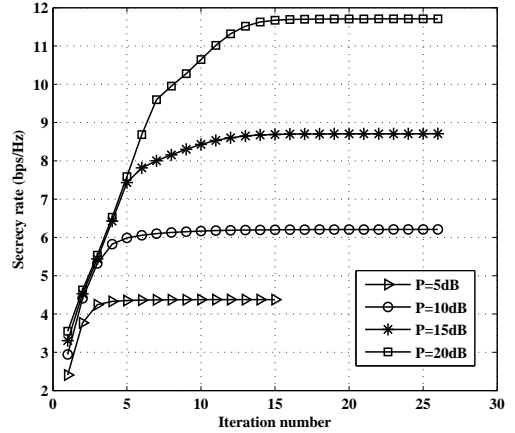}
\caption{DC algorithm: Convergence of the secrecy rate}\label{Convergence_Conf}
\end{center}
\end{figure}
First, we evaluate the convergence of the DC algorithm. Especially, we are concerned about whether the primal constraint (\ref{relax.a}) is violated by our approximation. Setting $\tau_{ms}$ as 2 bps/Hz, Fig.\,\ref{Convergence_Multi} shows the convergence of the multicast rate in the iteration with different transmit power. ${{\bf{\tilde Q}}_{c,0}}$ and ${{\bf{\tilde Q}}_{a,0}}$ are both initiated as $\bf{0}$. The algorithm stops iterating when the difference between two successive values of ${\bar{R}}(\tau _{ms})$ returned by the algorithm is less than or equal to $10^{-4}$. One can observe that, the multicast rates ultimately converge to our predefined multicast rate with a limited number of iterations in all tested transmit powers. This observation indicates the efficacy of TSE in approximating the multicast rate. Then we also plot the achieved secrecy rates and the approximated secrecy rates in Fig.\,\ref{Convergence_Conf}. The general observation of Fig.\,\ref{Convergence_Multi} is also applicable to Fig.\,\ref{Convergence_Conf}.

\begin{figure}[!t]
\begin{center}
\includegraphics[width=3in]{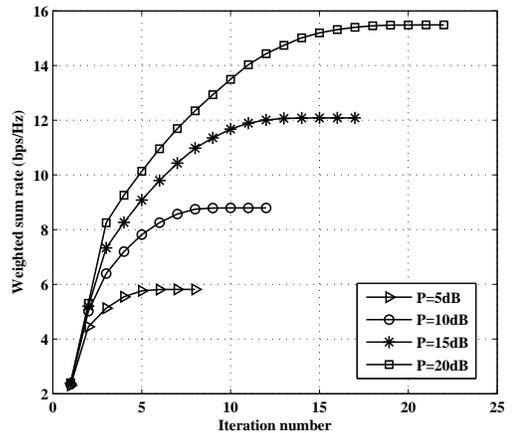}
\caption{AO algorithm: Convergence of the weighted sum rate}\label{Convergence_AO}
\end{center}
\end{figure}
The convergence results of the AO algorithm are presented in Fig.\,\ref{Convergence_AO}. In Fig.\,\ref{Convergence_AO}, we set $\lambda_c=1$ to seek the sum-rate point. ${\bf{Q}}_c^0$ and ${\bf{Q}}_a^0$ are both initialized as $(P/(2N_t)){\bf{I}}_{N_t}$. The algorithm stops iterating when the difference between two successive values of ${\bar{R}}(\lambda _{c})$ is less than or equal to $10^{-4}$. As one can observe from Fig.\,\ref{Convergence_AO}, the achieved weighted sum rate is monotonically increasing and finally converges with a limited number of iterations in all tested transmit powers. In addition, we find out that the AN covariance matrix $\mathbf{Q}_a$ output by AO is no longer diagonal. This implies that the associated AN design is spatially selective rather than isotropic, which blocks the eavesdroppers much more effectively. One can also note that the increase in the weighted sum rate is particularly remarkable when the transmit power is high. After all, higher transmit power means that the transmitter can allocate more power to the confidential message transmission, while not compromising the multicast performance. The extra power allocated to the confidential message can be used to generate more interference at the eavesdropper and/or strengthen the signal reception at the intended receiver, whereby more remarkable improvement is observed.

\subsection{Performance Comparison}
In this subsection, we focus on two sorts of system configuration. The first one is the same as that in the last subsection. Besides, we consider another sort of system configuration: $N_t=N_{r,1}=4$, $N_{r,k}=5$ for all $k \in {\cal{K}}_e$, and $K=4$. Under the second system configuration, \emph{neither} Condition \ref{C1} \emph{nor} Condition \ref{C2} is satisfied.

\begin{figure}[!t]
\begin{center}
\includegraphics[width=3in]{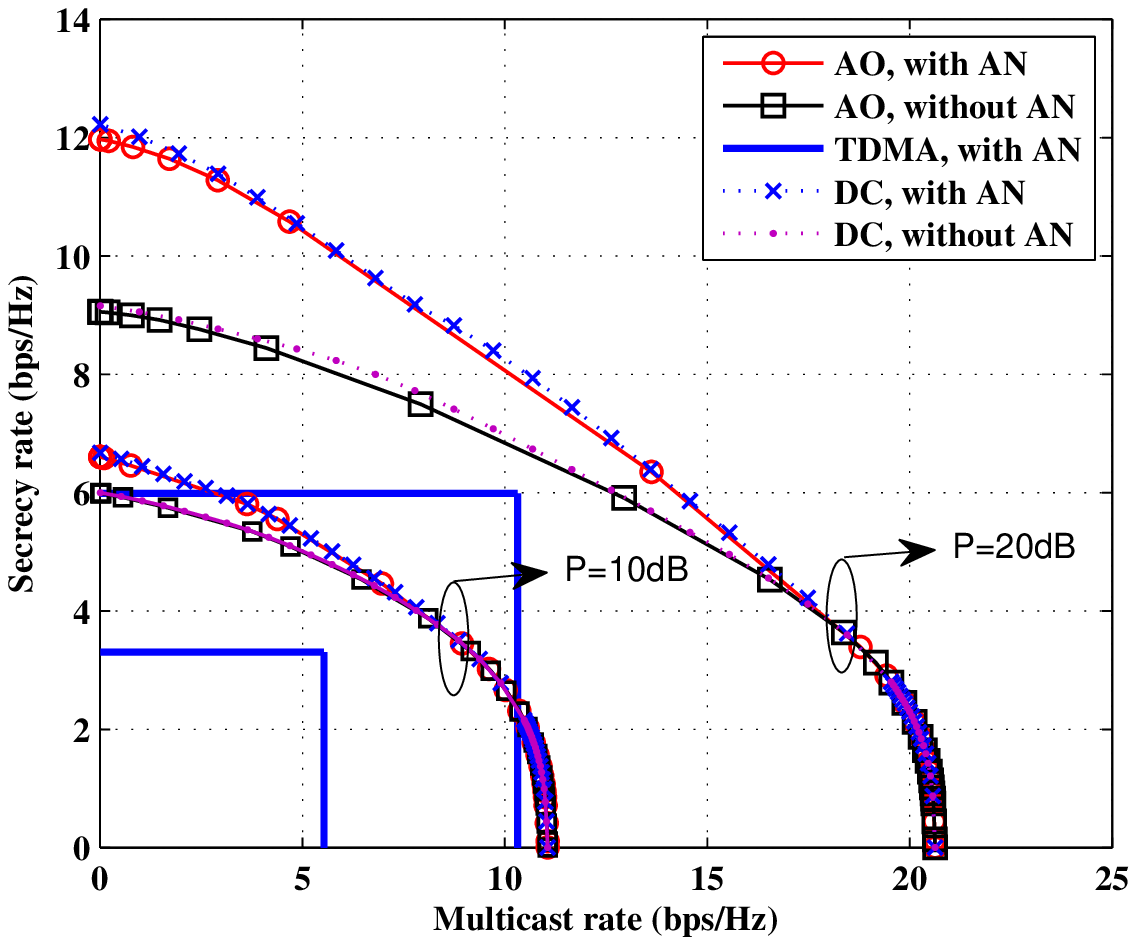}
\caption{Secrecy rate regions with and without AN.}\label{AO_DC1}
\end{center}
\end{figure}
First, we will show the secrecy rate regions achieved by the first system configuration. Overall results are shown in Fig.\,\ref{AO_DC1}, with $P$ set as $10$dB and $20$dB, respectively. Fig.\,\ref{AO_DC1} reveals two general trends. First, our AN-aided scheme achieves a secrecy rate region larger than the no-AN one. The striking gap indicates the efficacy of AN in expanding the secrecy rate region. However, the gap between these two strategies dramatically reduces when $R_0$ increases. This phenomenon agrees with our conjecture in Section \Rmnum{5}-A. The second observation is that our proposed strategies, though only attain a lower bound on $R_s({\left\{ {{\bf{H}}_k} \right\}_{k \in {\cal{K}}}},P)$, is sufficient to achieve significantly larger secrecy rate regions than the TDMA-based one. This observation also implies that PHY-SI is an effective approach to improve the spectral efficiency. Then let us compare the achievable performance of the two proposed scalarization methods. One can notice that the performance gap between these two methods is negligible in the tested system configuration, especially when $P=10$dB.

\begin{figure}[!t]
\begin{center}
\includegraphics[width=3in]{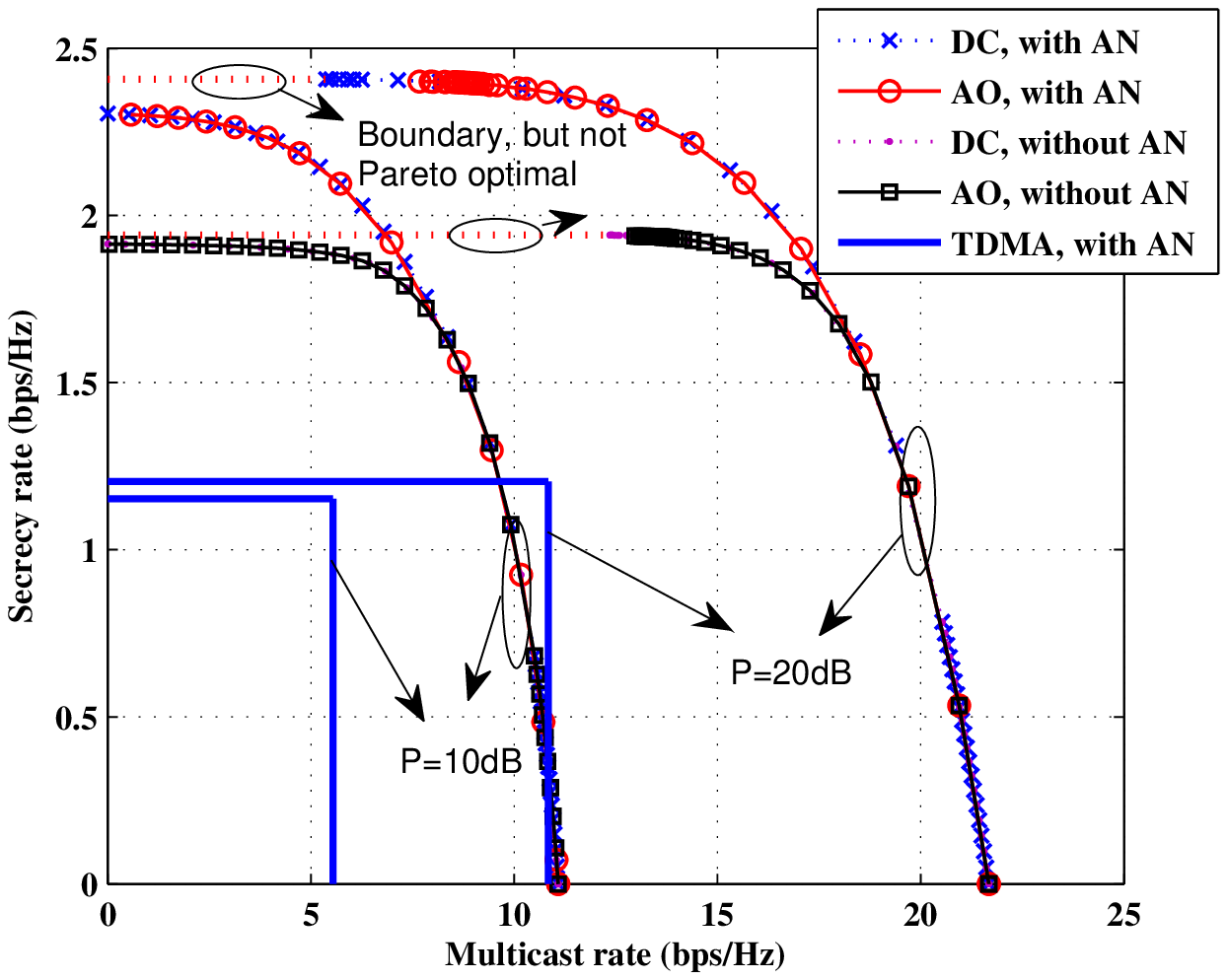}
\caption{Secrecy rate regions with and without AN.}\label{AO_DC2}
\end{center}
\end{figure}
Fig. \ref{AO_DC2} plots the secrecy rate regions achieved by the second system configuration. Still, the secrecy rate region with AN is larger than the one without AN and the one achieved by TDMA. Besides, we can observe two very interesting phenomena. First, when we increase the transmit power from 10dB to 20dB, the secrecy rate regions practically expand in the horizontal direction. That is, under the second system configuration, the increasing transmit power mainly contributes to the multicast message transmission, rather than the confidential message transmission. This can be interpreted from the transmit degree of freedom (d.o.f.). The total d.o.f. of unauthorized receivers is $\sum\nolimits_{k=2}^{K-1}{N_{r,k}}=15$, much higher than the transmit d.o.f. $N_t = 4$. The high d.o.f. at the unauthorized receivers leads to the d.o.f. bottleneck at the transmitter and thus compromises the overall secrecy performance. Second, one can notice that when $P=20$dB:

1) there exist some boundary points residing on a line, marked by the red dashed lines, that are not Pareto optimal to (\ref{op1}). Apparently, these points cannot be detected by the weighted-sum scalarization, but can be easily detected by the QoMS-based scalarization;

2) the QoMS-based scalarization detects more Pareto optimal points than the weighted-sum scalarization. This is attributed to the insensitivity of the weighted-sum scalarization to the points residing on an approximately horizontal boundary. To detect these boundary points, one has to precisely adjust the value of $\lambda_c$ to get different tangent points.

\subsection{Complexity Comparison}
\begin{table}[htbp]
  \centering
  \caption{Averaged running times (in secs.)}\label{running}
    \begin{tabular}{ccccccc}
    \toprule
    \multirow{2}[4]{*}{Method} &  \multicolumn{6}{c}{Power (dB)} \\
          & 0     & 4     & 8     & 12    & 16    & 20 \\
    \midrule
    DC algorithm & 6.07  & 8.89  & 12.91 & 17.35 & 21.18 & 30.84 \\
    AO algorithm & 7.57  & 11.58  &11.04 & 12.61 & 13.61 & 17.11 \\
    \bottomrule
    \end{tabular}%
\end{table}%
Finally, we tabulated the averaged running times of DC and AO for obtaining a boundary point in Table \ref{running} under the same setting as Fig.\,\ref{AO_DC1}. As seen, the DC algorithm runs faster than the AO algorithm when the transmit power is low. This phenomenon is consistent with our preceding analysis in Section \Rmnum{5}-B. However, at high transmit power, the DC algorithm scales nearly exponentially with $P$ and gradually spends more time converging in each iteration than the AO algorithm. This observation indicates that the two proposed scalarization methods might exhibit a performance-complexity tradeoff at high transmit power.

\section{Conclusion}
In this paper, we considered the AN-aided transmit design for multiuser MIMO broadcast channel with confidential service and multicast service. The transmit covariance matrices of confidential message, multicast message and AN were designed to maximize the achievable secrecy rate and achievable multicast rate simultaneously. To deal with this biobjective optimization problem, two different sorts of scalarization were introduced to transform this SRRM problem into a scalar optimization problem. In the QoMS-based scalarization, the scalar problem is an SRM problem with QoMS constraints, while in the weighted-sum scalarization, the scalar problem is a WSRM problem. DC and AO algorithms were utilized to solve the QoMS-constrained SRM problem and the WSRM problem, respectively. Both algorithms can converge to a stationary point of the primal problems. Further, we gave a detailed comparison between the two proposed scalarization methods. The comparison results indicated that at low transmit power, the QoMS-based scalarization is superior to the weighted-sum one in terms of achievable performance and computational complexity. On the other hand, at high transmit power, these two methods exhibit a tradeoff between achievable performance and computational complexity. Numerical results also confirmed the effectiveness of AN in expanding the secrecy rate region.

As a future direction, it would be interesting to analyze the robust service integration scheme to combat the possible CSI uncertainties caused by channel aging, and to take into account some application-specific requirements in 5G wireless communication system, e.g., the mobility of terminals and the overhead in CSI acquisition.

\appendix
\subsection{Proof of Proposition \ref{P1}}\label{DC_Appendix}
First, we claim that problem (\ref{relax}) has a following interesting property provided that Condition \ref{C1} or Condition \ref{C2} is satisfied.
\begin{property}\label{property1}
The maximum objective value of problem (\ref{relax}), ${R}({\tau_{ms}})$, is obtained only when the equality in (\ref{relax.a}) holds.
\end{property}
\begin{IEEEproof}
The proof of Property \ref{property1} can be accomplished by contradiction. Assume that the maximum value of problem (\ref{relax}) is obtained at the solution $({{\bf{\hat Q}}_0},{{\bf{\hat Q}}_c},{{\bf{\hat Q}}_a})$ and the equality in (\ref{relax.a}) does not hold, i.e., $\mathop {\min }\limits_{k \in \cal{K}} \log \lvert {{\bf{I}} + {{( {{\bf{I}} + {{\bf{H}}_k}({{\bf{\hat Q}}_c}+{{\bf{\hat Q}}_a}){\bf{H}}_k^H} )}^{ - 1}}{{\bf{H}}_k}{{\bf{\hat Q}}_0}{\bf{H}}_k^H} \rvert>\tau_{ms}$.

Our next step is to construct a new solution $({{\bf{\bar Q}}_0},{{\bf{\bar Q}}_c},{{\bf{\bar Q}}_a})$ from $({{\bf{\hat Q}}_0},{{\bf{\hat Q}}_c},{{\bf{\hat Q}}_a})$, which achieves a larger objective value and satisfies the constraint (\ref{relax.a}) with equality. Let us first elaborate upon the construction method under Condition \ref{C1}.

\subsubsection{Case for Condition \ref{C1}}Specifically, we multiply ${{\bf{\hat Q}}_0}$ by a scaling factor $\xi\;(0 < \xi < 1)$, add a positive semidefinite (PSD) matrix ${\bf{E}}=\rho{\bf{I}} - \rho{\bf{H}}_1^H{({{\bf{H}}_1}{\bf{H}}_1^H)^{ - 1}}{{\bf{H}}_1}$ to ${{\bf{\hat Q}}_a}$ and keep ${{\bf{\hat Q}}_c}$ constant, i.e., ${{\bf{\bar Q}}_0}=\xi{\bf{\hat Q}}_0$, ${{\bf{\bar Q}}_a}={{\bf{\hat Q}}_a}+\bf{E}$ and ${{\bf{\bar Q}}_c}={\bf{\hat Q}}_c$, where the coefficient $\rho$ controls the power of ${\bf{E}}$. Note that $\bf{E}$ is the orthogonal complement projector of ${\bf{H}}_1^H$, and its existence is guaranteed by Condition \ref{C1}. To keep the total transmit power constant, the coefficient $\rho$ should be chosen to satisfy
$(1 - \xi ){\rm{Tr}}({{{\bf{\hat Q}}}_0}) = {\rm{Tr}}({\bf{E}}) = \rho ({N_t} - {N_{r,1}})$, that is, $\rho  = \frac{{(1 - \xi ){\rm{Tr}}({{{\bf{\hat Q}}}_0})}}{{{N_t} - {N_{r,1}}}}$. To proceed, we need the following lemma.

\begin{lemma}[\cite{weingarten2006the}]\label{lem4}
For matrices ${\bf{A}},{\bf{\Delta }} \succeq {\bf{0}}$ and ${\bf{B}} \succ {\bf{0}}$, the following inequality hold:
\begin{equation}\label{lemma1}
\frac{{\left| {{\bf{A}} + {\bf{B}}} \right|}}{{\left| {\bf{B}} \right|}} \ge \frac{{\left| {{\bf{A}} + {\bf{B}} + {\bf{\Delta }}} \right|}}{{\left| {{\bf{B}} + {\bf{\Delta }}} \right|}}.
\end{equation}
\end{lemma}

Then, by applying Lemma \ref{lem4}, one can obtain
\begin{align}\label{CF2}
C&_{m,k}({{\bf{\hat Q}}_0},{{\bf{\hat Q}}_c},{{\bf{\hat Q}}_a})\nonumber\\
&=\log \lvert {{\bf{I}} + {{( {{\bf{I}} + {{\bf{H}}_k}({{\bf{\hat Q}}_c}+{{\bf{\hat Q}}_a}){\bf{H}}_k^H} )}^{ - 1}}{{\bf{H}}_k}{{\bf{\hat Q}}_0}{\bf{H}}_k^H} \rvert \nonumber\\
&> \log \lvert {{\bf{I}} + {{( {{\bf{I}} + {{\bf{H}}_k}({{\bf{\bar Q}}_c}+{{\bf{\bar Q}}_a}){\bf{H}}_k^H} )}^{ - 1}}{{\bf{H}}_k}{{\bf{\bar Q}}_0}{\bf{H}}_k^H} \rvert \nonumber\\
&=C_{m,k}({{\bf{\bar Q}}_0},{{\bf{\bar Q}}_c},{{\bf{\bar Q}}_a})
\end{align}
for any $k \in {\cal{K}}$. Thus, by adjusting the value of $\xi$, the equality in (\ref{relax}) could be achieved.

To proceed, we will show that a larger objective value could always be achieved by $({{\bf{\bar Q}}_0},{{\bf{\bar Q}}_c},{{\bf{\bar Q}}_a})$. By reapplying Lemma \ref{lem4}, it is easy to get
\begin{align}\label{CF3}
C&_{e,k}({{\bf{\bar Q}}_c},{{\bf{\bar Q}}_a})\nonumber\\
&=\log \lvert{\bf{I}} + {({\bf{I}} + {{\bf{H}}_k}{{{\bf{\bar Q}}}_a}{\bf{H}}_k^H)^{ - 1}}{{\bf{H}}_k}{{{\bf{\bar Q}}}_c}{\bf{H}}_k^H \rvert \nonumber\\
&= \log \lvert {{\bf{I}} + {{( {{\bf{I}} + {{\bf{H}}_k}({{\bf{\hat Q}}_a}+{\bf{E}}){\bf{H}}_k^H} )}^{ - 1}}{{\bf{H}}_k}{{\bf{\hat Q}}_c}{\bf{H}}_k^H} \rvert \nonumber\\
&< \log \lvert{\bf{I}} + {({\bf{I}} + {{\bf{H}}_k}{{{\bf{\hat Q}}}_a}{\bf{H}}_k^H)^{ - 1}}{{\bf{H}}_k}{{{\bf{\hat Q}}}_c}{\bf{H}}_k^H \rvert, \nonumber\\
&=C_{e,k}({{\bf{\hat Q}}_c},{{\bf{\hat Q}}_a}), \forall k \in {\cal{K}}_e.
\end{align}
Meanwhile, due to ${{\bf{H}}_1}{\bf{E}}{\bf{H}}_1^H=\bf{0}$, it is easy to see
\begin{equation}\label{CF4}
C_b({{\bf{\bar Q}}_c},{{\bf{\bar Q}}_a})=C_b({{\bf{\hat Q}}_c},{{\bf{\hat Q}}_a}).
\end{equation}

Combining (\ref{CF3}) with (\ref{CF4}), we obtain
\begin{align}\label{CF5}
C_b({{\bf{\bar Q}}_c}&,{{\bf{\bar Q}}_a})-\mathop {\max}\limits_{k \in {{\cal K}_e}}C_{e,k}({{\bf{\bar Q}}_c},{{\bf{\bar Q}}_a})\nonumber\\
&>C_b({{\bf{\hat Q}}_c},{{\bf{\hat Q}}_a})-\mathop {\max}\limits_{k \in {{\cal K}_e}}C_{e,k}({{\bf{\hat Q}}_c},{{\bf{\hat Q}}_a}),
\end{align}
i.e., a larger objective value can be found with $({{\bf{\bar Q}}_0},{{\bf{\bar Q}}_c},{{\bf{\bar Q}}_a})$. This fact is contrary to the primal assumption.

\subsubsection{Case for Condition \ref{C2}}The only difference between the proof for Condition \ref{C1} and Condition \ref{C2} lies in the construction method of $({{\bf{\bar Q}}_0},{{\bf{\bar Q}}_c},{{\bf{\bar Q}}_a})$. To begin with, let us first define a matrix ${{\bf{H}}_{ua}} \buildrel \Delta \over = {[{\bf{H}}_2^H,{\bf{H}}_3^H, \cdots ,{\bf{H}}_K^H]^H} \in {{\mathbb{C}}^{\sum\nolimits_{k \in {{\cal K}_e}} {{N_{r,k}}} \times {N_t}}}$, which stacks all of the unauthorized receivers' channel matrices. Then, we multiply ${{\bf{\hat Q}}_0}$ by a scaling factor $\xi\;(0 < \xi < 1)$, add a PSD matrix ${\bf{E}}=\rho{\bf{I}} - \rho{\bf{H}}_{ua}^H{({{\bf{H}}_{ua}}{\bf{H}}_{ua}^H)^{ - 1}}{{\bf{H}}_{ua}}$ to ${{\bf{\hat Q}}_c}$ and keep ${{\bf{\hat Q}}_a}$ constant, i.e., ${{\bf{\bar Q}}_0}=\xi{\bf{\hat Q}}_0$, ${{\bf{\bar Q}}_c}={{\bf{\hat Q}}_c}+\bf{E}$ and ${{\bf{\bar Q}}_a}={\bf{\hat Q}}_a$, where the coefficient $\rho$ controls the power of ${\bf{E}}$. $\bf{E}$ is the orthogonal complement projector of ${\bf{H}}_{ua}^H$, the existence of which is guaranteed by Condition \ref{C2}. The coefficient $\rho$ should be chosen to satisfy $\rho  = \frac{{(1 - \xi ){\rm{Tr}}({{{\bf{\hat Q}}}_0})}}{{N_t} - \sum\nolimits_{k \in {{\cal K}_e}} {{N_{r,k}}}}$ to keep the total transmit power constant.

Again, by exploiting Lemma \ref{lem4} and carrying out some matrix manipulations, one can verify that $({{\bf{\bar Q}}_0},{{\bf{\bar Q}}_c},{{\bf{\bar Q}}_a})$ can achieve a larger objective value than $({{\bf{\hat Q}}_0},{{\bf{\hat Q}}_c},{{\bf{\hat Q}}_a})$ with the constraint (\ref{relax.a}) active. This fact contradicts the primal assumption.

Summarizing the conclusions drawn from the two cases above, we have accomplished the proof of Property \ref{property1}.
\end{IEEEproof}

Property \ref{property1} makes the proof of \cite[Theorem 1]{mei2016secrecy} fully applicable to the proposition here. The remaining parts of the proof can be found in \cite{mei2016secrecy} and are omitted here for simplicity.

\subsection{Proof of Proposition \ref{KKTConv}}\label{AO_Appendix}
Firstly, we introduce slack variables $\alpha$ and $\beta$ to reexpress (\ref{op3}) as
\begin{subequations}\label{op10}
\begin{align}
\nonumber \mathop {\max}\limits_{{{\bf{Q}}_0},{{\bf{Q}}_c},{{\bf{Q}}_a},\alpha,\beta}&\; \lambda_c (C_b - \beta)+ \alpha\\
\text{s.t.}\quad& {C_{e,k}} \le \beta, \forall k \in {\cal{K}}_e,\label{op10a}\\
& {C_{m,k}} \ge \alpha, \forall k \in {\cal{K}},\label{op10b}\\
&\nonumber\text{(\ref{op7a})-(\ref{op7b}) are satisfied.}
\end{align}
\end{subequations}
Equivalently, it suffices to prove that every limit point $({\bf{\tilde Q}}_0,{\bf{\tilde Q}}_c,{\bf{\tilde Q}}_a)$ of the iterates generated by the AO algorithm, together with $\tilde \alpha  = \mathop {\min }\limits_{k \in {\cal{K}}} {C_{m,k}}({\bf{\tilde Q}}_0,{\bf{\tilde Q}}_c,{\bf{\tilde Q}}_a)$ and $\tilde \beta  = \mathop {\max }\limits_{k \in {\cal{K}}_e} {C_{e,k}}({\bf{\tilde Q}}_0,{\bf{\tilde Q}}_c,{\bf{\tilde Q}}_a)$ is a KKT point of (\ref{op10}).
\begin{figure*}[t]
\normalsize
\setcounter{MYtempeqncnt}{\value{equation}}
\setcounter{equation}{45}
\begin{subequations}\label{kkt7}
\begin{align}
&\lambda_c{\nabla _{{{\bf{Q}}_c}}}{\varphi _b}({{{\bf{\tilde Q}}}_c},{{{\bf{\tilde Q}}}_a},{{{\bf{\tilde S}}}_1})- \sum\limits_{k \in {\cal{K}}_e} {{\rho _k}{\nabla _{{{\bf{Q}}_c}}}{\varphi _{e,k}}({{{\bf{\tilde Q}}}_c},{{{\bf{\tilde Q}}}_a},{{{\bf{\tilde S}}}_k})}+\sum\limits_{k \in {\cal{K}}} {{\mu _k}{\nabla _{{{\bf{Q}}_c}}}{\varphi _{m,k}}({{{\bf{\tilde Q}}}_0},{{{\bf{\tilde Q}}}_c},{{{\bf{\tilde Q}}}_a},{{{\bf{\tilde U}}}_k})}- \gamma {\bf{I}} + {\bf{C}} = {\bf{0}}, \label{kkt7a}\\
&\lambda_c{\nabla _{{{\bf{Q}}_a}}}{\varphi _b}({{{\bf{\tilde Q}}}_c},{{{\bf{\tilde Q}}}_a},{{{\bf{\tilde S}}}_1})- \sum\limits_{k \in {\cal{K}}_e} {{\rho _k}{\nabla _{{{\bf{Q}}_a}}}{\varphi _{e,k}}({{{\bf{\tilde Q}}}_c},{{{\bf{\tilde Q}}}_a},{{{\bf{\tilde S}}}_k})}+\sum\limits_{k \in {\cal{K}}} {{\mu _k}{\nabla _{{{\bf{Q}}_a}}}{\varphi _{m,k}}({{{\bf{\tilde Q}}}_0},{{{\bf{\tilde Q}}}_c},{{{\bf{\tilde Q}}}_a},{{{\bf{\tilde U}}}_k})}- \gamma {\bf{I}} + {\bf{A}} = {\bf{0}}, \label{kkt7b}\\
&\sum\limits_{k \in {\cal{K}}} {{\mu _k}{\nabla _{{{\bf{Q}}_0}}}{\varphi _{m,k}}({{{\bf{\tilde Q}}}_0},{{{\bf{\tilde Q}}}_c},{{{\bf{\tilde Q}}}_a},{{{\bf{\tilde U}}}_k})}- \gamma {\bf{I}} + {\bf{B}}  = {\bf{0}}, \label{kkt7c}\\
&{\varphi _{e,k}}({{{\bf{\tilde Q}}}_c},{{{\bf{\tilde Q}}}_a},{{{\bf{\tilde S}}}_k}) \le \tilde \beta, \forall k \in {\cal{K}}_e \label{kkt7d}\\
&{\rho _k}({\varphi _{e,k}}({{{\bf{\tilde Q}}}_c},{{{\bf{\tilde Q}}}_a},{{{\bf{\tilde S}}}_k}) - \tilde \beta) = 0, \forall k \in {\cal{K}}_e \label{kkt7e}\\
&{\varphi _{m,k}}({{{\bf{\tilde Q}}}_0},{{{\bf{\tilde Q}}}_c},{{{\bf{\tilde Q}}}_a},{{{\bf{\tilde U}}}_k}) \ge \tilde \alpha , \forall k \in {\cal{K}}\label{kkt7f}\\
&{\mu _k}({\varphi _{m,k}}({{{\bf{\tilde Q}}}_0},{{{\bf{\tilde Q}}}_c},{{{\bf{\tilde Q}}}_a},{{{\bf{\tilde U}}}_k}) - \tilde \alpha) = 0, \forall k \in {\cal{K}}\label{kkt7g} \\
&\sum\nolimits_{k = 1}^K {{\rho _k}}  = 1, \label{kkt7h}\\
&\sum\nolimits_{k = 1}^K {{\mu _k}}  = 1, \label{kkt7i}\\
&{\bf{A}} \succeq {\bf{0}},{\bf{B}} \succeq {\bf{0}},{\bf{C}} \succeq {\bf{0}},\label{kkt7j}\\
&\gamma  \ge 0, {{\rho _k}} \ge 0, \forall k \in {\cal{K}}_e, {{\mu _k}} \ge 0, \forall k \in {\cal{K}},\label{kkt7k}\\
&{\rm{Tr}}({{{\bf{\tilde Q}}}_0} +{{{\bf{\tilde Q}}}_c} + {{{\bf{\tilde Q}}}_a}) \le P,{{{\bf{\tilde Q}}}_0} \succeq {\bf{0}},{{{\bf{\tilde Q}}}_c} \succeq {\bf{0}},{{{\bf{\tilde Q}}}_a} \succeq {\bf{0}},\label{kkt7l}\\
&\gamma ({\rm{Tr}}({{{\bf{\tilde Q}}}_0} +{{{\bf{\tilde Q}}}_c} + {{{\bf{\tilde Q}}}_a}) - P) = 0,\label{kkt7m}\\
&{\rm{Tr}}({\bf{B}}{{{\bf{\tilde Q}}}_0}) = 0, {\rm{Tr}}({\bf{C}}{{{\bf{\tilde Q}}}_c}) = 0,{\rm{Tr}}({\bf{A}}{{{\bf{\tilde Q}}}_a}) = 0,\label{kkt7n}
\end{align}
\end{subequations}
\hrulefill
\setcounter{equation}{48}
\begin{subequations}\label{kkt10}
\begin{align}
&\lambda_c{\nabla _{{{\bf{Q}}_c}}}{C_b}({{{\bf{\tilde Q}}}_c},{{{\bf{\tilde Q}}}_a})- \sum\limits_{k \in {\cal{K}}_e} {{\rho _k}{\nabla _{{{\bf{Q}}_c}}}{C_{e,k}}({{{\bf{\tilde Q}}}_c},{{{\bf{\tilde Q}}}_a})}- \gamma {\bf{I}} + {\bf{C}}+\sum\limits_{k \in {\cal{K}}} {{\mu _k}{\nabla _{{{\bf{Q}}_c}}}{C_{m,k}}({{{\bf{\tilde Q}}}_0},{{{\bf{\tilde Q}}}_c},{{{\bf{\tilde Q}}}_a})} = {\bf{0}}, \label{kkt10a}\\
&\lambda_c{\nabla _{{{\bf{Q}}_a}}}{C_b}({{{\bf{\tilde Q}}}_c},{{{\bf{\tilde Q}}}_a})- \sum\limits_{k \in {\cal{K}}_e} {{\rho _k}{\nabla _{{{\bf{Q}}_a}}}{C_{e,k}}({{{\bf{\tilde Q}}}_c},{{{\bf{\tilde Q}}}_a})}- \gamma {\bf{I}} + {\bf{A}}+\sum\limits_{k \in {\cal{K}}} {{\mu _k}{\nabla _{{{\bf{Q}}_a}}}{C_{m,k}}({{{\bf{\tilde Q}}}_0},{{{\bf{\tilde Q}}}_c},{{{\bf{\tilde Q}}}_a})}= {\bf{0}}, \label{kkt10b}\\
&\sum\limits_{k \in {\cal{K}}}{\mu _k}{\nabla _{{{\bf{Q}}_0}}}{C_{m,k}}({{{\bf{\tilde Q}}}_0},{{{\bf{\tilde Q}}}_c},{{{\bf{\tilde Q}}}_a})- \gamma {\bf{I}} + {\bf{B}} = {\bf{0}}, \label{kkt10c}\\
&{C_{e,k}}({{{\bf{\tilde Q}}}_c},{{{\bf{\tilde Q}}}_a}) \le \tilde \beta, \forall k \in {\cal{K}}_e \label{kkt10d}\\
&{\rho _k}({C_{e,k}}({{{\bf{\tilde Q}}}_c},{{{\bf{\tilde Q}}}_a}) - \tilde \beta) = 0, \forall k \in {\cal{K}}_e \label{kkt10e}\\
&{C_{m,k}}({{{\bf{\tilde Q}}}_0},{{{\bf{\tilde Q}}}_c},{{{\bf{\tilde Q}}}_a}) \ge \tilde \alpha , \forall k \in {\cal{K}}\label{kkt10f}\\
&{\mu _k}({C_{m,k}}({{{\bf{\tilde Q}}}_0},{{{\bf{\tilde Q}}}_c},{{{\bf{\tilde Q}}}_a}) - \tilde \alpha) = 0, \forall k \in {\cal{K}}\label{kkt10g}\\
&\text{(\ref{kkt7h})-(\ref{kkt7n}) satisfied.}\label{kkt10h}
\end{align}
\end{subequations}
\vspace*{-12pt}
\hrulefill
\setcounter{equation}{\value{MYtempeqncnt}}
\end{figure*}

Due to the compactness of $({\bf{Q}}_0,{\bf{Q}}_c,{\bf{Q}}_a)$, there must exist a subsequence, denoted by ${\{ ({\bf{Q}}_0^{{n_l}},{\bf{Q}}_c^{{n_l}},{\bf{Q}}_a^{{n_l}},\{ {\bf{S}}_k^{{n_l}}\} _{k = 1}^K,\{ {\bf{U}}_k^{{n_l}}\} _{k = 1}^K)\}_l}$, such that ${\{ ({\bf{Q}}_0^{{n_l}},{\bf{Q}}_c^{{n_l}},{\bf{Q}}_a^{{n_l}},\{ {\bf{S}}_k^{{n_l}}\} _{k = 1}^K,\{ {\bf{U}}_k^{{n_l}}\} _{k = 1}^K)\}_l}$ converges to a limit point $({\bf{\tilde Q}}_0,{\bf{\tilde Q}}_c,{\bf{\tilde Q}}_a,\{{\bf{\tilde S}}_k\} _{k = 1}^K,\{{\bf{\tilde U}}_k\} _{k = 1}^K)$ as $l \to \infty$. Next, our proof is composed of two steps. First, we will show that the limit point $({\bf{\tilde Q}}_0,{\bf{\tilde Q}}_c,{\bf{\tilde Q}}_a,\{{\bf{\tilde S}}_k\} _{k = 1}^K,\{{\bf{\tilde U}}_k\} _{k = 1}^K)$ satisfy the following properties.
\begin{subequations}\label{kkt1}
\begin{align}
&{\bf{\tilde S}}_1 = \arg \;\mathop {\max }\limits_{{{\bf{S}}_1} \succeq {\bf{0}}} {\varphi _b}({{{\bf{\tilde Q}}}_c},{{{\bf{\tilde Q}}}_a},{{\bf{S}}_1}), \label{kkt1a}\\
&{\bf{\tilde S}}_k = \arg \;\mathop {\min }\limits_{{{\bf{S}}_k} \succeq {\bf{0}}} {\varphi _{e,k}}({{{\bf{\tilde Q}}}_c},{{{\bf{\tilde Q}}}_a},{{\bf{S}}_k}),\forall k \in {\cal{K}}_e \label{kkt1b}\\
&{\bf{\tilde U}}_k = \arg \;\mathop {\max }\limits_{{{\bf{U}}_k} \succeq {\bf{0}}} {\varphi _{m,k}}({{{\bf{\tilde Q}}}_0},{{{\bf{\tilde Q}}}_c},{{{\bf{\tilde Q}}}_a},{{\bf{U}}_k}),\forall k \in {\cal{K}} \label{kkt1c}\\
\nonumber &({{{\bf{\tilde Q}}}_0},{{{\bf{\tilde Q}}}_c},{{{\bf{\tilde Q}}}_a})= \\
&\quad\arg\;\mathop {\max}\limits_{({{\bf{Q}}_0},{{\bf{Q}}_c},{{\bf{Q}}_a}) \in {\cal{F}}} \;f({{\bf{Q}}_0},{{\bf{Q}}_c},{{\bf{Q}}_a},\{{\bf{\tilde S}}_k\} _{k = 1}^K,\{{\bf{\tilde U}}_k\} _{k = 1}^K). \label{kkt1d}
\end{align}
\end{subequations}
Second, we will check the KKT conditions of problems (\ref{kkt1a})-(\ref{kkt1d}) to build a bridge between (\ref{kkt1}) and the KKT conditions of problem (\ref{op10}).

\textbf{Step 1.} By noting that
\begin{subequations}\label{kkt2}
\begin{align}
&{\bf{S}}_1^{{n_l}} = \arg \;\mathop {\max }\limits_{{{\bf{S}}_1} \succeq {\bf{0}}} {\varphi _b}({\bf{Q}}_c^{{n_l} - 1},{\bf{Q}}_a^{{n_l} - 1},{{\bf{S}}_1}),\label{kkt2a}\\
&{\bf{S}}_k^{{n_l}} = \arg \;\mathop {\min }\limits_{{{\bf{S}}_k} \succeq {\bf{0}}} {\varphi _{e,k}}({\bf{Q}}_c^{{n_l} - 1},{\bf{Q}}_a^{{n_l} - 1},{{\bf{S}}_k}), \forall k \in {\cal{K}}_e \label{kkt2b}\\
&{\bf{U}}_k^{{n_l}} = \arg \;\mathop {\max }\limits_{{{\bf{U}}_k} \succeq {\bf{0}}} {\varphi _{m,k}}({\bf{Q}}_0^{{n_l} - 1},{\bf{Q}}_c^{{n_l} - 1},{\bf{Q}}_a^{{n_l} - 1},{{\bf{U}}_k}), \forall k \in {\cal{K}} \label{kkt2c} \\
\nonumber &({\bf{Q}}_0^{{n_l}},{\bf{Q}}_c^{{n_l}},{\bf{Q}}_a^{{n_l}}) = \\
&\quad \arg\;\mathop {\max}\limits_{({{\bf{Q}}_0},{{\bf{Q}}_c},{{\bf{Q}}_a}) \in {\cal{F}}} \;f({{\bf{Q}}_0},{{\bf{Q}}_c},{{\bf{Q}}_a},\{ {\bf{S}}_k^{{n_l}}\} _{k = 1}^K,\{ {\bf{U}}_k^{{n_l}}\} _{k = 1}^K),\label{kkt2d}
\end{align}
\end{subequations}
we have
\begin{subequations}\label{kkt3}
\begin{align}
&{\varphi _b}({\bf{Q}}_c^{{n_l} - 1},{\bf{Q}}_a^{{n_l} - 1},{\bf{S}}_1^{{n_l}}) \ge {\varphi _b}({\bf{Q}}_c^{{n_l} - 1},{\bf{Q}}_a^{{n_l} - 1},{{\bf{S}}_1}), \forall {{\bf{S}}_1} \succeq \bf{0} \label{kkt3a}\\
\nonumber&{\varphi _{e,k}}({\bf{Q}}_c^{{n_l} - 1},{\bf{Q}}_a^{{n_l} - 1},{\bf{S}}_k^{{n_l}}) \le {\varphi _{e,k}}({\bf{Q}}_c^{{n_l} - 1},{\bf{Q}}_a^{{n_l} - 1},{{\bf{S}}_k}),\\
&\forall {{\bf{S}}_k} \succeq \mathbf{0}, \forall k \in {\cal{K}}_e \label{kkt3b}\\
\nonumber&{\varphi _{m,k}}({\bf{Q}}_0^{{n_l} - 1},{\bf{Q}}_c^{{n_l} - 1},{\bf{Q}}_a^{{n_l} - 1},{\bf{U}}_k^{{n_l}}) \ge {\varphi _{m,k}}({\bf{Q}}_0^{{n_l} - 1},{\bf{Q}}_c^{{n_l} - 1},\\
&{\bf{Q}}_a^{{n_l} - 1},{{\bf{U}}_k}),\forall {{\bf{U}}_k} \succeq \mathbf{0}, \forall k \in {\cal{K}} \label{kkt3c}
\end{align}
\end{subequations}
and
\begin{align}
&f({{\bf{Q}}_0},{{\bf{Q}}_c},{{\bf{Q}}_a},\{ {\bf{S}}_k^{{n_l}}\} _{k = 1}^K,\{ {\bf{U}}_k^{{n_l}}\} _{k = 1}^K) \nonumber\\
\le &f({\bf{Q}}_0^{{n_l}},{\bf{Q}}_c^{{n_l}},{\bf{Q}}_a^{{n_l}},\{ {\bf{S}}_k^{{n_l}}\} _{k = 1}^K,\{ {\bf{U}}_k^{{n_l}}\} _{k = 1}^K)\label{kkt4}\\
\le &f({{{\bf{\tilde Q}}}_0},{{{\bf{\tilde Q}}}_c},{{{\bf{\tilde Q}}}_a},\{ {{{\bf{\tilde S}}}_k}\} _{k = 1}^K), \{{{{\bf{\tilde U}}}_k}\} _{k = 1}^K), \forall ({{\bf{Q}}_0},{{\bf{Q}}_c},{{\bf{Q}}_a}) \in {\cal{F}},\nonumber
\end{align}
where the second inequality of (\ref{kkt4}) holds for the reason that AO algorithm yields non-descending objective values. Then letting $l \to \infty$ in (\ref{kkt3}) and (\ref{kkt4}) will lead to (\ref{kkt1a})-(\ref{kkt1d}).

\textbf{Step 2.} Then it follows from (\ref{kkt1a})-(\ref{kkt1d}) and the positive definiteness of $\{ {{{\bf{\tilde S}}}_k}\} _{k = 1}^K$ and $\{ {{{\bf{\tilde U}}}_k}\} _{k = 1}^K$ that
\begin{subequations}\label{kkt5}
\begin{align}
&{\nabla _{{{\bf{S}}_1}}}{\varphi _b}({{{\bf{\tilde Q}}}_c},{{{\bf{\tilde Q}}}_a},{{{\bf{\tilde S}}}_1})=\mathbf{0}, {{{\bf{\tilde S}}}_1} \succeq \mathbf{0}, \label{kkt5a}\\
&{\nabla _{{{\bf{S}}_k}}}{\varphi _{e,k}}({{{\bf{\tilde Q}}}_c},{{{\bf{\tilde Q}}}_a},{{{\bf{\tilde S}}}_k}) = {\bf{0}},{{{\bf{\tilde S}}}_k} \succeq {\bf{0}}, \forall k \in {\cal{K}}_e \label{kkt5b}\\
&{\nabla _{{{\bf{U}}_k}}}{\varphi _{m,k}}({{{\bf{\tilde Q}}}_0},{{{\bf{\tilde Q}}}_c},{{{\bf{\tilde Q}}}_a},{{{\bf{\tilde U}}}_k}) = {\bf{0}},{{{\bf{\tilde U}}}_k} \succeq {\bf{0}}, \forall k \in {\cal{K}}. \label{kkt5c}
\end{align}
\end{subequations}
By carrying out some matrix manipulations to (\ref{kkt5}), it is easy to obtain that
\begin{subequations}\label{kkt6}
\begin{align}
&{{{\bf{\tilde S}}}_1} = {({\bf{I}} + {{\bf{H}}_1}{{{\bf{\tilde Q}}}_a}{\bf{H}}_1^H)^{ - 1}} \succeq {\bf{0}},\\
&{{{\bf{\tilde S}}}_k} = {({\bf{I}} + {{\bf{H}}_{k}}({{{\bf{\tilde Q}}}_a} + {{{\bf{\tilde Q}}}_c}){\bf{H}}_{k}^H)^{ - 1}}\succeq {\bf{0}}, \forall k \in {\cal{K}}_e \label{kkt6b}\\
&{{{\bf{\tilde U}}}_k} = {({\bf{I}} + {{\bf{H}}_{k}}({{{\bf{\tilde Q}}}_a} + {{{\bf{\tilde Q}}}_c}){\bf{H}}_{k}^H)^{ - 1}}\succeq {\bf{0}}, \forall k \in {\cal{K}} \label{kkt6c}
\end{align}
\end{subequations}
Meanwhile, by introducing slack variables $\alpha$ and $\beta$, (\ref{kkt1d}) is shown to be equivalent to
\begin{subequations}\label{op11}
\begin{align}
\nonumber \mathop {\max}\limits_{{{\bf{Q}}_0},{{\bf{Q}}_c},{{\bf{Q}}_a},\alpha,\beta} &\lambda_c ({\varphi _b}({{\bf{Q}}_c},{{\bf{Q}}_a},{{{\bf{\tilde S}}}_1}) - \beta) + \alpha \\
\text{s.t.}\quad &{\varphi _{e,k}}({{\bf{Q}}_c},{{\bf{Q}}_a},{{{\bf{\tilde S}}}_k}) \le \beta , \forall k \in {\cal{K}}_e,\label{op11a}\\
&{\varphi _{m,k}}({{\bf{Q}}_0},{{\bf{Q}}_c},{{\bf{Q}}_a},{{{\bf{\tilde U}}}_k}) \ge \alpha , \forall k \in {\cal{K}},\label{op11b}\\
&({{\bf{Q}}_0},{{\bf{Q}}_c},{{\bf{Q}}_a}) \in {\cal{F}}.
\end{align}
\end{subequations}
It is easy to see that $({{{\bf{\tilde Q}}}_0},{{{\bf{\tilde Q}}}_c},{{{\bf{\tilde Q}}}_a})$, together with $\tilde \beta  = \mathop {\max }\limits_{k \in {\cal{K}}_e} {\varphi_{e,k}}({\bf{\tilde Q}}_c,{\bf{\tilde Q}}_a,{{{\bf{\tilde S}}}_k})$ and $\tilde \alpha  = \mathop {\min }\limits_{k \in {\cal{K}}} {\varphi_{m,k}}({\bf{\tilde Q}}_0,{\bf{\tilde Q}}_c,{\bf{\tilde Q}}_a,{{{\bf{\tilde U}}}_k})$ is an optimal solution of problem (\ref{op4}). Consequently, $({{{\bf{\tilde Q}}}_0},{{{\bf{\tilde Q}}}_c},{{{\bf{\tilde Q}}}_a},\tilde \beta,\tilde \alpha)$ satisfy the KKT conditions of (\ref{op11}), shown in \addtocounter{equation}{1}(\ref{kkt7}) at the top of last page. In (\ref{kkt7}), $\left(\{{\rho _k}\} _{k \in {\cal{K}}_e},\{{\mu _k}\} _{k \in {\cal{K}}},\gamma,{\bf{A}},{\bf{B}},{\bf{C}}\right)$ are all dual variables pertaining to the constraints in (\ref{op11}).

To proceed, by applying Danskin's theorem \cite{Bertsekas1999}, one can verify the following equalities must hold.
\begin{subequations}\label{kkt8}
\begin{align}
&{\nabla _{{\bf{Q}}_c}}{C_b}({{{\bf{\tilde Q}}}_c},{{{\bf{\tilde Q}}}_a}) = {\nabla _{{\bf{Q}}_c}}{\varphi _b}({{{\bf{\tilde Q}}}_c},{{{\bf{\tilde Q}}}_a},{{{\bf{\tilde S}}}_1}),\label{kkt8a}\\
&{\nabla _{{\bf{Q}}_a}}{C_b}({{{\bf{\tilde Q}}}_c},{{{\bf{\tilde Q}}}_a}) = {\nabla _{{\bf{Q}}_a}}{\varphi _b}({{{\bf{\tilde Q}}}_c},{{{\bf{\tilde Q}}}_a},{{{\bf{\tilde S}}}_1}),\label{kkt8b}\\
&{\nabla _{{\bf{Q}}_c}}{C_{e,k}}({{{\bf{\tilde Q}}}_c},{{{\bf{\tilde Q}}}_a}) = {\nabla _{{\bf{Q}}_c}}{\varphi _{e,k}}({{{\bf{\tilde Q}}}_c},{{{\bf{\tilde Q}}}_a},{{{\bf{\tilde S}}}_k}),  \label{kkt8c}\\
&{\nabla _{{\bf{Q}}_a}}{C_{e,k}}({{{\bf{\tilde Q}}}_c},{{{\bf{\tilde Q}}}_a}) = {\nabla _{{\bf{Q}}_a}}{\varphi _{e,k}}({{{\bf{\tilde Q}}}_c},{{{\bf{\tilde Q}}}_a},{{{\bf{\tilde S}}}_k}), \label{kkt8d}\\
&{\nabla _{{\bf{Q}}_c}}{C_{m,k}}({{{\bf{\tilde Q}}}_0},{{{\bf{\tilde Q}}}_c},{{{\bf{\tilde Q}}}_a}) = {\nabla _{{\bf{Q}}_c}}{\varphi _{m,k}}({{{\bf{\tilde Q}}}_0},{{{\bf{\tilde Q}}}_c},{{{\bf{\tilde Q}}}_a},{{{\bf{\tilde U}}}_k}),\label{kkt8e}\\
&{\nabla _{{\bf{Q}}_a}}{C_{m,k}}({{{\bf{\tilde Q}}}_0},{{{\bf{\tilde Q}}}_c},{{{\bf{\tilde Q}}}_a}) = {\nabla _{{\bf{Q}}_a}}{\varphi _{m,k}}({{{\bf{\tilde Q}}}_0},{{{\bf{\tilde Q}}}_c},{{{\bf{\tilde Q}}}_a},{{{\bf{\tilde U}}}_k}), \label{kkt8f}\\
&{\nabla _{{\bf{Q}}_0}}{C_{m,k}}({{{\bf{\tilde Q}}}_0},{{{\bf{\tilde Q}}}_c},{{{\bf{\tilde Q}}}_a}) = {\nabla _{{\bf{Q}}_0}}{\varphi _{m,k}}({{{\bf{\tilde Q}}}_0},{{{\bf{\tilde Q}}}_c},{{{\bf{\tilde Q}}}_a},{{{\bf{\tilde U}}}_k}).\label{kkt8g}
\end{align}
\end{subequations}

Then substituting (\ref{kkt6b}) and (\ref{kkt6c}) into ${\varphi _{e,k}}({{{\bf{\tilde Q}}}_c},{{{\bf{\tilde Q}}}_a},{{{\bf{\tilde S}}}_k})$ and ${\varphi _{m,k}}({{{\bf{\tilde Q}}}_0},{{{\bf{\tilde Q}}}_c},{{{\bf{\tilde Q}}}_a},{{{\bf{\tilde U}}}_k})$, one can obtain
\begin{subequations}\label{kkt9}
\begin{align}
&{C_{e,k}}({{{\bf{\tilde Q}}}_c},{{{\bf{\tilde Q}}}_a},{{\tilde p}_m}) = {\varphi _{e,k}}({{{\bf{\tilde Q}}}_c},{{{\bf{\tilde Q}}}_a},{{\tilde p}_m},{{{\bf{\tilde S}}}_k}), \forall k \in {\cal{K}}_e\\
&{C_{m,k}}({{{\bf{\tilde Q}}}_0},{{{\bf{\tilde Q}}}_c},{{{\bf{\tilde Q}}}_a}) = {\varphi _{e,k}}({{{\bf{\tilde Q}}}_0},{{{\bf{\tilde Q}}}_c},{{{\bf{\tilde Q}}}_a},{{{\bf{\tilde U}}}_k}), \forall k \in {\cal{K}}
\end{align}
\end{subequations}

Finally, by plunging (\ref{kkt8}) and (\ref{kkt9}) into (\ref{kkt7}), we obtain \addtocounter{equation}{1}(\ref{kkt10}), shown at the top of last page. Remarkably, (\ref{kkt10a})-(\ref{kkt10h}) are KKT conditions of the WSRM problem (\ref{op10}). This fact completes the proof.

\bibliography{PHYSI_MIMO}

\begin{thebibliography}{10}
\providecommand{\url}[1]{#1}
\csname url@samestyle\endcsname
\providecommand{\newblock}{\relax}
\providecommand{\bibinfo}[2]{#2}
\providecommand{\BIBentrySTDinterwordspacing}{\spaceskip=0pt\relax}
\providecommand{\BIBentryALTinterwordstretchfactor}{4}
\providecommand{\BIBentryALTinterwordspacing}{\spaceskip=\fontdimen2\font plus
\BIBentryALTinterwordstretchfactor\fontdimen3\font minus
  \fontdimen4\font\relax}
\providecommand{\BIBforeignlanguage}[2]{{%
\expandafter\ifx\csname l@#1\endcsname\relax
\typeout{** WARNING: IEEEtran.bst: No hyphenation pattern has been}%
\typeout{** loaded for the language `#1'. Using the pattern for}%
\typeout{** the default language instead.}%
\else
\language=\csname l@#1\endcsname
\fi
#2}}
\providecommand{\BIBdecl}{\relax}
\BIBdecl

\bibitem{andrew2014what}
J.~G. Andrews, S.~Buzzi, W.~Choi, S.~V. Hanly, A.~Lozano, A.~C.~K. Soong, and
  J.~C. Zhang, ``What will {5G} be?'' \emph{{IEEE} J. Sel. Areas Commun.},
  vol.~32, no.~6, pp. 1065--1082, Jun. 2014.

\bibitem{shiu2011physical}
Y.-S. Shiu, S.~Y. Chang, H.-C. Wu, S.~C.-H. Huang, and H.-H. Chen, ``Physical
  layer security in wireless networks: a tutorial,'' \emph{{IEEE} Wireless
  Commun.}, vol.~18, no.~2, pp. 66--74, Apr. 2011.

\bibitem{he2013wireless}
\BIBentryALTinterwordspacing
B.~He, X.~Zhou, and T.~D. Abhayapala, ``Wireless physical layer security with
  imperfect channel state information: A survey,'' Jun. 2013. [Online].
  Available: \url{http://arxiv.org/abs/1307.4146}
\BIBentrySTDinterwordspacing

\bibitem{hong2013enhancing}
Y.-W.~P. Hong, P.-C. Lan, and C.-C.~J. Kuo, ``Enhancing physical-layer secrecy
  in multiantenna wireless systems: An overview of signal processing
  approaches,'' \emph{{IEEE} Signal Process. Mag.}, vol.~30, no.~5, pp. 29--40,
  Sep. 2013.

\bibitem{mukherjee2014principles}
A.~Mukherjee, S.~A. Fakoorian, J.~Huang, A.~L. Swindlehurst \emph{et~al.},
  ``Principles of physical layer security in multiuser wireless networks: A
  survey,'' \emph{{IEEE} Commun. Surveys Tuts.}, vol.~16, no.~3, pp.
  1550--1573, Aug. 2014.

\bibitem{liu2016physical}
Y.~Liu, H.-H. Chen, and L.~Wang, ``Physical layer security for next generation
  wireless networks: Theories, technologies, and challenges,'' \emph{{IEEE}
  Commun. Surveys Tuts.}, vol.~19, no.~1, pp. 347--376, 2017.

\bibitem{liu2009secrecy}
R.~Liu and H.~V. Poor, ``Secrecy capacity region of a multi-antenna {G}aussian
  broadcast channel with confidential messages,'' \emph{{IEEE} Trans. Inf.
  Theory}, vol.~55, no.~3, pp. 1235--1249, Mar. 2009.

\bibitem{liu2010multiple}
R.~Liu, T.~Liu, H.~V. Poor, and S.~Shamai, ``Multiple-input multiple-output
  {G}aussian broadcast channels with confidential messages,'' \emph{{IEEE}
  Trans. Inf. Theory}, vol.~56, no.~9, pp. 4215--4227, Sep. 2010.

\bibitem{fakoorian2013on}
S.~A.~A. Fakoorian and A.~L. Swindlehurst, ``On the optimality of linear
  precoding for secrecy in the {MIMO} broadcast channel,'' \emph{{IEEE} J. Sel.
  Areas Commun.}, vol.~31, no.~9, pp. 1701--1713, Sep. 2013.

\bibitem{park2016weighted}
D.~Park, ``Weighted sum rate maximization of {MIMO} broadcast and interference
  channels with confidential messages,'' \emph{{IEEE} Trans. Wireless Commun.},
  vol.~15, no.~3, pp. 1742--1753, Mar. 2016.

\bibitem{csiszar1978broadcast}
I.~Csisz{\'a}r and J.~K{\"o}rner, ``Broadcast channels with confidential
  messages,'' \emph{{IEEE} Trans. Inf. Theory}, vol.~24, no.~3, pp. 339--348,
  May 1978.

\bibitem{Hung2010Multiple}
H.~D. Ly, T.~Liu, and Y.~Liang, ``Multiple-input multiple-output {G}aussian
  broadcast channels with common and confidential messages,'' \emph{{IEEE}
  Trans. Inf. Theory}, vol.~56, no.~11, pp. 5477--5487, Oct. 2010.

\bibitem{ekrem2012capacity}
E.~Ekrem and S.~Ulukus, ``Capacity region of gaussian {MIMO} broadcast channels
  with common and confidential messages,'' \emph{{IEEE} Trans. Inf. Theory},
  vol.~58, no.~9, pp. 5669--5680, Sep. 2012.

\bibitem{liu2010mimo}
R.~Liu, T.~Liu, H.~V. Poor, and S.~Shamai, ``{MIMO} {G}aussian broadcast
  channels with confidential and common messages,'' in \emph{Proc. {IEEE} Int.
  Symp. Inf. Theory ({ISIT}'2010)}, Austin, TX, Jun. 2010, pp. 2578--2582.

\bibitem{liu2013new}
R.~Liu, T.~Liu, H.~V. Poor, and S.~Shamai~(Shitz), ``New results on
  multiple-input multiple-output broadcast channels with confidential
  messages,'' \emph{{IEEE} Trans. Inf. Theory}, vol.~59, no.~3, pp. 1346--1358,
  Mar. 2013.

\bibitem{wyrembelski2011service}
R.~F. Wyrembelski and H.~Boche, ``Service integration in multiantenna
  bidirectional relay networks: Public and confidential messages,'' in
  \emph{Proc. {IEEE} Global Communication Conf. Workshops}, Houston, TX, Dec.
  2011, pp. 884--888.

\bibitem{Wyrembelski2012Physical}
R.~Wyrembelski and H.~Boche, ``Physical layer integration of private, common,
  and confidential messages in bidirectional relay networks,'' \emph{{IEEE}
  Trans. Wireless Commun.}, vol.~11, no.~9, pp. 3170--3179, Sep. 2012.

\bibitem{Schaefer2014Physical}
R.~Schaefer and H.~Boche, ``Physical layer service integration in wireless
  networks: Signal processing challenges,'' \emph{{IEEE} Signal Process. Mag.},
  vol.~31, no.~3, pp. 147--156, Apr. 2014.

\bibitem{mei2016secrecy}
W.~Mei, Z.~Chen, and J.~Fang, ``Secrecy capacity region maximization in
  {G}aussian {MISO} channels with integrated services,'' \emph{{IEEE} Signal
  Process. Lett.}, vol.~23, no.~8, pp. 1146--1150, Jul. 2016.

\bibitem{mei1512}
W.~Mei, L.~Li, Z.~Chen, and C.~Huang, ``Artificial-noise aided transmit design
  for multi-user {MISO} systems with integrated services,'' in \emph{Proc.
  {IEEE} Global Conf. Signal Info. Process. (GlobalSIP)}, Orlando, FL, Dec.
  2015, pp. 1382--1386.

\bibitem{mei2016robust}
W.~Mei, Z.~Chen, and C.~Huang, ``Robust artificial-noise aided transmit design
  for multi-user {MISO} systems with integrated services,'' in \emph{Proc.
  {IEEE} ICASSP}, Shanghai, Mar. 2016, pp. 3856--3860.

\bibitem{mei2016artificial}
W.~Mei, L.~Li, Z.~Chen, and C.~Huang, ``Artificial-noise aided transmit design
  for outage constrained service integration,'' in \emph{Proc. {IEEE} Int.
  Conf. Commun.}, Kuala Lumpur, Malaysia, May 2016, pp. 1--7.

\bibitem{mei2016GSVD}
W.~Mei, Z.~Chen, and J.~Fang, ``{GSVD}-based precoding in {MIMO} systems with
  integrated services,'' \emph{{IEEE} Signal Process. Lett.}, vol.~23, no.~11,
  pp. 1528--1532, Sep. 2016.

\bibitem{li2013transmit}
Q.~Li, M.~Hong, H.-T. Wai, Y.-F. Liu, W.-K. Ma, and Z.-Q. Luo, ``Transmit
  solutions for {MIMO} wiretap channels using alternating optimization,''
  \emph{{IEEE} J. Sel. Areas Commun.}, vol.~31, no.~9, pp. 1714--1727, Sep.
  2013.

\bibitem{li2013spatially}
Q.~Li and W.-K. Ma, ``Spatially selective artificial-noise aided transmit
  optimization for {MISO} multi-{E}ves secrecy rate maximization,''
  \emph{{IEEE} Trans. Signal Process.}, vol.~61, no.~10, pp. 2704--2717, May
  2013.

\bibitem{chu2015robust}
Z.~Chu, K.~Cumanan, Z.~Ding, M.~Johnston, and S.~Y. Le~Goff, ``Robust outage
  secrecy rate optimizations for a {MIMO} secrecy channel,'' \emph{{IEEE}
  Wireless Commun. Lett.}, vol.~4, no.~1, pp. 86--89, Feb. 2015.

\bibitem{chu2016secrecy}
Z.~Chu, H.~Xing, M.~Johnston, and S.~Y. Le~Goff, ``Secrecy rate optimizations
  for a {MISO} secrecy channel with multiple multiantenna eavesdroppers,''
  \emph{{IEEE} Trans. Wireless Commun.}, vol.~15, no.~1, pp. 283--297, Jan.
  2016.

\bibitem{zheng2015multi}
T.-X. Zheng, H.-M. Wang, J.~Yuan, D.~Towsley, and M.~H. Lee, ``Multi-antenna
  transmission with artificial noise against randomly distributed
  eavesdroppers,'' \emph{{IEEE} Trans. Commun.}, vol.~63, no.~11, pp.
  4347--4362, Nov. 2015.

\bibitem{NIPS2009}
G.~R. Lanckriet and B.~K. Sriperumbudur, ``On the convergence of the
  concave-convex procedure,'' in \emph{Proc. Advances Neural Inf. Process.
  Syst.}, 2009, pp. 1759--1767.

\bibitem{fang2016precoding}
B.~Fang, Z.~Qian, W.~Shao, and W.~Zhong, ``Precoding and artificial noise
  design for cognitive {MIMOME} wiretap channels,'' \emph{{IEEE} Trans. Veh.
  Technol.}, vol.~65, no.~8, pp. 6753--6758, Aug. 2016.

\bibitem{chu2015secrecy}
Z.~Chu, K.~Cumanan, Z.~Ding, M.~Johnston, and S.~Y. Le~Goff, ``Secrecy rate
  optimizations for a {MIMO} secrecy channel with a cooperative jammer,''
  \emph{{IEEE} Trans. Veh. Technol.}, vol.~64, no.~5, pp. 1833--1847, May 2015.

\bibitem{yang2013optimal}
J.~Yang, I.-M. Kim, and D.~I. Kim, ``Optimal cooperative jamming for multiuser
  broadcast channel with multiple eavesdroppers,'' \emph{{IEEE} Trans. Wireless
  Commun.}, vol.~12, no.~6, pp. 2840--2852, Jun. 2013.

\bibitem{wu2013physical}
S.~X. Wu, W.-K. Ma, and A.~M.-C. So, ``Physical-layer multicasting by
  stochastic transmit beamforming and {A}lamouti space-time coding,''
  \emph{{IEEE} Trans. Signal Process.}, vol.~61, no.~17, pp. 4230--4245, Sep.
  2013.

\bibitem{zhu2012precoder}
H.~Zhu, N.~Prasad, and S.~Rangarajan, ``Precoder design for physical layer
  multicasting,'' \emph{{IEEE} Trans. Signal Process.}, vol.~60, no.~11, pp.
  5932--5947, Nov. 2012.

\bibitem{lee2013a}
W.~Lee, H.~Park, H.~B. Kong, J.~S. Kwak, and I.~Lee, ``A new beamforming design
  for multicast systems,'' \emph{{IEEE} Trans. Veh. Technol.}, vol.~62, no.~8,
  pp. 4093--4097, Oct. 2013.

\bibitem{du2013optimum}
B.~Du, Y.~Jiang, X.~Xu, and X.~Dai, ``Optimum beamforming for {MIMO}
  multicasting,'' \emph{EURASIP J. Adv. Signal Process.}, vol. 2013, no. 121,
  pp. 1--15, Dec. 2013.

\bibitem{Boyd2011CVX}
\BIBentryALTinterwordspacing
M.~Grant and S.~Boyd, ``{CVX}: {M}atlab software for disciplined convex
  programming,'' Apr. 2011. [Online]. Available: \url{http://cvxr.com/cvx}
\BIBentrySTDinterwordspacing

\bibitem{boyd2009convex}
S.~Boyd and L.~Vandenberghe, \emph{Convex optimization}.\hskip 1em plus 0.5em
  minus 0.4em\relax Cambridge, UK: Cambridge university press, 2009.

\bibitem{marler2004survey}
R.~T. Marler and J.~S. Arora, ``Survey of multi-objective optimization methods
  for engineering,'' \emph{Structural Multidisciplinary Optim.}, vol.~26,
  no.~6, pp. 369--395, Apr. 2004.

\bibitem{luo2010semi}
Z.-Q. Luo, W.-K. Ma, A.~M.-C. So, Y.~Ye, and S.~Zhang, ``Semidefinite
  relaxation of quadratic optimization problems,'' \emph{{IEEE} Signal Process.
  Mag.}, vol.~27, no.~3, pp. 20--34, May 2010.

\bibitem{cumanan2014secrecy}
K.~Cumanan, Z.~Ding, B.~Sharif, G.~Y. Tian, and K.~K. Leung, ``Secrecy rate
  optimizations for a {MIMO} secrecy channel with a multiple-antenna
  eavesdropper,'' \emph{{IEEE} Trans. Veh. Technol.}, vol.~63, no.~4, pp.
  1678--1690, May 2014.

\bibitem{ben2001lectures}
A.~Ben-Tal and A.~Nemirovski, \emph{Lectures on modern convex optimization:
  analysis, algorithms, and engineering applications}.\hskip 1em plus 0.5em
  minus 0.4em\relax Philadelphia, PA, USA: SIAM, 2001, vol.~2.

\bibitem{weingarten2006the}
H.~Weingarten, Y.~Steinberg, and S.~Shamai~(Shitz), ``The capacity region of
  the {G}aussian multiple-input multiple-output broadcast channel,''
  \emph{{IEEE} Trans. Inf. Theory}, vol.~52, no.~9, pp. 3936--3964, Sep. 2006.

\bibitem{Bertsekas1999}
D.~Bertsekas, \emph{Nonlinear Programming}, 2nd~ed.\hskip 1em plus 0.5em minus
  0.4em\relax Belmont, MA, USA: Athena Scientific, 1999.

\end{thebibliography}
\bibliographystyle{IEEEtran}

\end{document}